\pdfoutput=1
\documentclass[pra,letterpaper,aps,10pt,superscriptaddress,twocolumn,floatfix,showpacs]{revtex4-2}
\usepackage[T1]{fontenc}
\usepackage[utf8]{inputenc}
\usepackage[english]{babel} 
\usepackage{graphicx}
\usepackage{amsmath,amssymb,amsfonts}
\usepackage{braket,xcolor,upgreek}
\usepackage[colorlinks, linkcolor=blue, citecolor=blue, urlcolor=blue, breaklinks=true]{hyperref}
\usepackage{orcidlink}
\usepackage{microtype}
\usepackage{bbm}
\usepackage{color}
\usepackage{listings}
\usepackage{upquote}
\usepackage{tikz}
\usetikzlibrary{arrows.meta,positioning,calc,fit,shapes.geometric}

\newcommand{\fref}[1]{Fig.~\ref{#1}}
\newcommand{\secref}[1]{Sec.~\ref{#1}}
\newcommand{\appref}[1]{Appendix~\ref{#1}}
\newcommand{\coderef}[1]{Code sample~\ref{#1}}

\renewcommand{\eqref}[1]{Eq.~(\ref{#1})}
\newcommand{\eqsref}[1]{Eqs.~(\ref{#1})}

\graphicspath{{figures/}}

\usepackage{listings}
\definecolor{ghbg}{HTML}{F6F8FA}       
\definecolor{ghfg}{HTML}{1F2328}       
\definecolor{ghcomment}{HTML}{6E7781}  
\definecolor{ghkeyword}{HTML}{CF222E}  
\definecolor{ghentity}{HTML}{8250DF}   
\definecolor{ghstring}{HTML}{0A3069}   
\definecolor{ghfunction}{HTML}{0550AE}  
\definecolor{codered}{rgb}{0.796,0.235,0.2}  
\definecolor{backcolour}{HTML}{F6F8FA} 

\newcommand{\lstsub}[1]{\raisebox{0pt}[0pt][0pt]{\textsubscript{#1}}}
\newcommand{\lstsup}[1]{\raisebox{0pt}[0pt][0pt]{\textsuperscript{#1}}}
\makeatletter
\lst@AddToHook{EveryPar}{\strut}
\makeatother

\lstdefinelanguage{Julia}{
    keywords=[1]{abstract, break, catch, const, continue, do, else, elseif, begin, end, export, false, for, function, if, import, let, local, macro, module, quote, return, struct, true, try, using, while, where},
    keywords=[2]{Bool, Char, Dict, Float64, Int, String, Symbol, Matrix, Real, Number, Complex, Array, AbstractArray, Vector, Set, Tuple, NamedTuple, QuantumObject, QuantumObjectEvolution, FockSpace, FockBasis, NLevelSpace, NLevelBasis, SLH, Transition, Destroy, Create, LazyKet, System, ODEProblem, Tsit5, EnzymeVJP, BacksolveAdjoint},
    keywords=[3]{dummy_start, versioninfo, Qobj, QobjEvo, tensor, basis, fock, destroy, qeye, sigmax, sigmay, sigmaz, sigmam, sigmap, range, sesolve, mesolve, mcsolve, ssesolve, smesolve, sqrt, sin, cos, steadystate, steadystate_fourier, dfd_mesolve, dsf_mesolve, coherent, sum, size, dropdims, H_dsf, c_ops_dsf, e_ops_dsf, cu, CuVector, wigner, normalize, addprocs, SlurmManager, set_num_threads, Lattice, mapreduce, multisite_operator, DissipativeIsing, Val, ntuple, EnsembleSplitThreads, rmprocs, workers, gradient, my_f_mesolve, real, expect, BacksolveAdjoint, EnzymeVJP, liouvillian, exp, conj, abs, adjoint, coef_a, coef_ac, hamiltonian, jump_operator, lindblad, scattering, cascade, concatenate, feedback, to_numeric, correlation_matrix, effective_output_mode, effective_input_mode, coupling_input, coupling_output, coupling_delay_in, coupling_delay_out, coupling_matrix, substitute, variable, meanfield, complete, mtkcompile, dagger, dense, static_operator, fockstate, nlevelstate, coherentstate, initial_values, get_solution, transition, simplify, eigen, eigenvectors, solve_mode_evolution, master, master_dynamic, solve, one, fill, zeros, length, max, merge, enumerate, dummy_end},
    alsoletter={@},
    keywords=[4]{@qnumbers, @register_symbolic, @named, @variables, @parameters, @independent_variables},
    sensitive=true,  
    morecomment=[l]\#,%
    morecomment=[n]{\#=}{=\#},%
    morestring=[s]{"}{"},%
    literate={
        {ᵥ₁}{{\lstsub{v1}}}2 {ᵥ₂}{{\lstsub{v2}}}2
        {₁₂}{{\lstsub{12}}}2 {ᵤᵥ}{{\lstsub{uv}}}2
        {ᵢₙ}{{\lstsub{in}}}2 {ₚₕ}{{\lstsub{ph}}}2
        {ᵤ}{{\lstsub{u}}}1 {ᵥ}{{\lstsub{v}}}1 {ₛ}{{\lstsub{s}}}1 {ₐ}{{\lstsub{a}}}1
        {ᵣ}{{\lstsub{R}}}1 {ₗ}{{\lstsub{L}}}1 {ⱼ}{{\lstsub{j}}}1 {ₑ}{{\lstsub{e}}}1
        {ₜ}{{\lstsub{t}}}1 {ᵢ}{{\lstsub{i}}}1 {ₙ}{{\lstsub{n}}}1 {ₚ}{{\lstsub{p}}}1
        {ₕ}{{\lstsub{h}}}1
        {₀}{{\lstsub{0}}}1 {₁}{{\lstsub{1}}}1 {₂}{{\lstsub{2}}}1 {₃}{{\lstsub{3}}}1
        {¹}{{\lstsup{(1)}}}1 {²}{{\lstsup{(2)}}}1
        {√}{{{\color{codered}$\surd$}}}{1}
        {\\}{{{\color{codered}\lstum@backslash}}}{1} {\{}{{{\color{codered}\{}}}{1}
        {\}}{{{\color{codered}\}}}}{1} 
        {\%}{{{\color{codered}\%}}}{1} {&}{{{\color{codered}\&}}}{1}
        {+}{{{\color{codered}+}}}{1}
        {*}{{{\color{codered}*}}}{1}
        {/}{{{\color{codered}/}}}{1}
        {'}{{{\color{codered}\textquotesingle}}}{1}
        {=}{{{\color{codered}=}}}{1}
        {<}{{{\color{codered}<}}}{1}
        {=>}{{{\color{codered}=>}}}{1} {|>}{{{\color{codered}|>}}}{1}
        {==}{{{\color{codered}==}}}{1}
        {>}{{{\color{codered}>}}}{1} {?}{{{\color{codered}?}}}{1}
        {julia>}{{{\color{codered}julia>}}}{1}
        {^}{{{\color{codered}\textasciicircum}}}{1} {|}{{{\color{codered}|}}}{1}
        {~}{{{\color{codered}\textasciitilde{}}}}{1}
        {×}{{{\color{codered}$\times$}}}{1} {⋅}{{{\color{codered}$\cdot$}}}{1}
        {π}{{$\pi$}}1
        {ω}{{$\omega$}}1
        {α}{{$\alpha$}}1
        {β}{{$\beta$}}1
        {σ}{{$\sigma$}}1
        {ψ}{{$\psi$}}1
        {κ}{{$\kappa$}}1
        {γ}{{$\gamma$}}1
        {Δ}{{$\Delta$}}1
        {∆}{{$\Delta$}}1
        {φ}{{$\phi$}}1
        {∘}{{$\circ$}}1
        {⊗}{{{\color{codered}$\otimes$}}}{1}
        {▷}{{{\color{codered}$\triangleright$}}}{1}
        {⊞}{{{\color{codered}$\boxplus$}}}{1}
        {τ}{{$\tau$}}1
        {ρ}{{$\rho$}}1
        {ϕ}{{$\phi$}}1
    }
}

\lstdefinestyle{mainstyle}{
    language=Julia,
    backgroundcolor=\color{ghbg},   
    commentstyle=\color{ghcomment}\itshape,  
    keywordstyle=[1]\color{ghkeyword},  
    keywordstyle=[2]\color{ghfunction}, 
    keywordstyle=[3]\color{ghfunction}, 
    keywordstyle=[4]\color{ghentity},   
    identifierstyle=\color{ghfg},
    stringstyle=\color{ghstring},
    basicstyle=\ttfamily\footnotesize\color{ghfg},
    columns=fullflexible,
    upquote=true,
    breakatwhitespace=false,         
    breaklines=true,                 
    captionpos=b,                    
    keepspaces=true,                 
    showspaces=false,                
    showstringspaces=false,
    showtabs=false,                  
    tabsize=2,
}

\lstdefinestyle{tablestyle}{
    language=Julia,
    backgroundcolor=\color{backcolour},
    keywordstyle=[2]\color{black},  
    keywordstyle=[3]\color{black},  
    basicstyle=\ttfamily\footnotesize,
    columns=fullflexible,
    upquote=true,
    breakatwhitespace=false,         
    breaklines=true,                 
    captionpos=b,                    
    keepspaces=true,                 
    numbers=left,                    
    numbersep=5pt,                  
    showspaces=false,                
    showstringspaces=false,
    showtabs=false,                  
    tabsize=2,
}

\newcommand{\code}[1]{%
  \begingroup
  \setlength{\fboxsep}{1pt} 
  \colorbox{backcolour}{\lstinline[basicstyle=\ttfamily]{#1}}%
  \endgroup
}
\newcommand{\codesymbol}[1]{%
  \begingroup
  \setlength{\fboxsep}{1pt} 
  \colorbox{backcolour}{#1}%
  \endgroup
}

\begin{document}

\title{\texorpdfstring{QuantumInputOutput.jl: A Julia framework for input--output theory with quantum pulses and the SLH formalism}{QuantumInputOutput.jl: A Julia framework for the input--output theory with quantum pulses and the SLH formalism}}

\author{Christoph Hotter\,\orcidlink{0009-0003-3854-0264}}
\affiliation{Center for Hybrid Quantum Networks (Hy-Q), Niels Bohr Institute, University of Copenhagen, Jagtvej 155A, Copenhagen DK-2200, Denmark} 
\affiliation{Department of Physics, MIT-Harvard Center for Ultracold Atoms and Research Laboratory of Electronics, Massachusetts Institute of Technology, Cambridge, Massachusetts 02139, USA} 

\author{Orjan Ameye\,\orcidlink{0000-0001-7082-7300}}
\affiliation{Department of Physics, University of Konstanz, D-78457 Konstanz, Germany} 

\author{Klaus Mølmer\,\orcidlink{0000-0002-2372-869X}}
\affiliation{Center for Hybrid Quantum Networks (Hy-Q), Niels Bohr Institute, University of Copenhagen, Jagtvej 155A, Copenhagen DK-2200, Denmark} 

\date{\today}

\begin{abstract}
The interaction of traveling photonic pulses with quantum systems is a fundamental process for many quantum technologies such as quantum communication, control, and measurement protocols. The so-called SLH formalism, named for the triple of operators that parametrize the dynamics, in combination with virtual emission and absorption cavities provides a general toolbox to describe such processes. However, modeling open quantum systems driven by propagating quantum pulses becomes cumbersome once several components, channels, and temporal modes are involved. Here we present \texttt{QuantumInputOutput.jl}, an open-source framework that automates this derivation process: first, symbolic expressions for the Hamiltonian and Lindblad operators are derived using predefined SLH rules and symbolic quantum algebra. Next, the obtained operators are converted to (time-dependent) numerical matrices or are used to obtain higher-order mean-field equations. The numerical expressions are presented in a form immediately suitable for solving the time evolution of the master equation and analyzing the output modes via correlation functions. Besides its usefulness for numerical simulations, the framework also provides a convenient tool to derive symbolic expressions for one-dimensional quantum systems such as multilevel emitters in cavities or waveguides.
\end{abstract}

\maketitle
 
\section{Introduction}
\label{sec:intro}

The input--output theory is a cornerstone of quantum optics~\cite{gardiner1985input}. It links the internal dynamics of a localized quantum system to the fields that enter and leave it. Experimentally accessible quantities, such as transmitted, reflected, or emitted radiation, thereby become computable directly from the system operators.
The framework of cascaded systems shows how the output field from one open quantum system can coherently drive another, establishing a master-equation description for directional quantum networks~\cite{gardiner1993driving,carmichael1993trajectory, gardiner2004quantum}.
This viewpoint has shaped the theory of open quantum systems, chiral light--matter interfaces~\cite{Pichler2015, Lodahl2017,combes2017slh}, waveguide QED and quantum networks~\cite{Sheremet2023}. 
These ideas have been used to describe several systems, e.g.\ collective emission in cascaded many-body systems~\cite{Stannigel2012, CardenasLopez2023,Kusmierek2023, liedl2024superradiant, Maffei2024}, giant-atom interactions~\cite{Kockum2014,Kockum2018,Kannan2020}, strongly interacting propagating photons~\cite{Fleischhauer2005eit, Nikoghosyan2010, Witthaut2012, Chang2014, caneva2015generalized}, coherent-feedback enhancement of squeezing~\cite{gough2009feedback}, and photon--atom gates~\cite{duan2004scalable,chan2022quantum,aqua2025atom}. A particularly useful tool to build more complex input--output networks is the SLH framework, where algebraic composition rules connect components represented by a \texttt{S}cattering matrix, \texttt{L}indblad operators, and a \texttt{H}amiltonian~\cite{gough2009series,gough2009hamiltonian,combes2017slh}.

An important class of input--output problems for modern quantum technologies concerns propagating quantum pulses. Photons are natural carriers of quantum information because they travel over long distances with weak decoherence~\cite{Kimble2008,koji2023repeaters}. 
The ability to generate, shape, scatter, interfere, detect, and analyze such photonic modes is therefore essential for quantum communication, sensing, and distributed quantum processing~\cite{brecht2015temporal,raymer2020temporal,fabre2020modes}. For a traveling quantum pulse, the temporal mode of a photon is part of the quantum state itself. The physically relevant object is not only an integrated photon number or a local system observable, but the quantum state distributed over one or several temporal modes of the emitted field.
This description has found applications in a variety of systems, e.g.\ Dicke superradiance~\cite{lemberger2021radiation}, information extraction from superradiant bursts~\cite{belliardo2026superradiantinfo}, nonlinear splitting and merging of few-photon wave packets~\cite{Yang2022,lund2023perfect}, the generation of photonic states from collective atomic states~\cite{gutman2024universal,tziperman2025nonlinear}, the two-color SUPER excitation of quantum dots~\cite{kerber2026super}, and the relation between the Jaynes--Cummings model and traveling light pulses~\cite{christiansen2024jc}.

\begin{figure}
    \centering 
    \includegraphics[width=0.95\columnwidth]{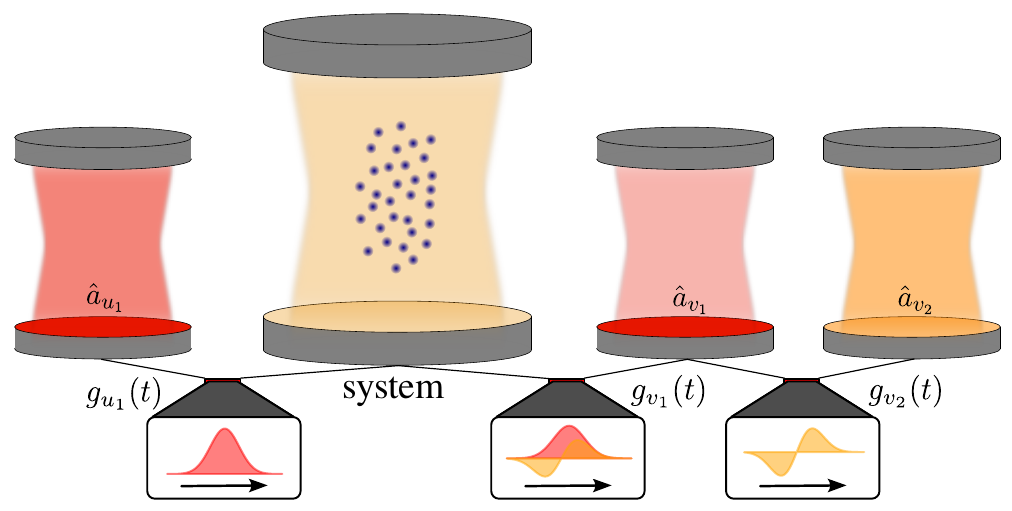}
    \caption{Input--output model with one virtual input cavity, a physical system, and two virtual output cavities. The field propagates along the cascade from left to right. The input cavity emits the mode $u_1(t)$, while the output cavities capture the modes $v_1(t)$ and $v_2(t)$, respectively.}
    \label{fig:io_model}
\end{figure}

Several frameworks address related waveguide-QED problems from complementary perspectives. Packages such as WaveguideQED.jl~\cite{bundgaard2025waveguideqed} and QwaveMPS~\cite{regidor2026qwavemps} use time-bin discretization or matrix product states to represent the propagating field explicitly. These approaches are powerful, in particular for non-Markovian feedback and spatially extended field states. However, the discretization of the continuum usually restricts them to smaller systems. The input--output theory with quantum pulses provides a finite-mode route through this continuum problem~\cite{kiilerich2019inputoutput,kiilerich2020quantum}. Instead of discretizing the propagating field, selected temporal wave packets are represented by virtual cavities with time-dependent couplings. A virtual cavity emits a prescribed temporal input mode, downstream virtual cavities absorb chosen output modes, and the physical system is placed between them in a cascaded master equation, see \fref{fig:io_model}. This construction keeps the Markovian continuum implicit while retaining a full quantum-state description of the system and the relevant pulse modes. 

Even in this finite-mode representation, deriving the effective Hamiltonian and Lindblad operators becomes laborious as components, channels, and temporal modes are added. The practical difficulty is not only algebraic: the network connectivity, the temporal modes used to represent the field, and the numerical treatment must be combined consistently. Other software packages address parts of this problem. QNET~\cite{tezak2012qhdl,qnet_github} composes general SLH networks symbolically but does not represent quantum pulse modes. The scattering module of QuTiP~\cite{Johansson2013,fischer2018scattering} computes the photons scattered into a waveguide by a system driven with classical pulses. 

In this paper we present \texttt{QuantumInputOutput.jl}~\cite{qio_github}, a Julia~\cite{bezanson2017julia} package for constructing quantum input--output networks with propagating pulse modes. It derives symbolic time-dependent Hamiltonians and Lindblad operators from user-defined SLH components and temporal pulse modes, converts them into numerical operators for master-equation simulations~\cite{Kraemer2018,QuantumToolbox2025}, or generates higher-order mean-field equations~\cite{Plankensteiner2022quantumcumulantsjl}. 
The symbolic results can be inspected, simplified, and used directly for analytic calculations. Because the derived models have Lindblad form, they can be passed to standard numerical master-equation solvers. In addition, the package provides tools for determining and analyzing relevant temporal output modes from correlation functions, transforming models to efficient interaction pictures~\cite{christiansen2023interactions}, and
incorporating propagation delays~\cite{christiansen2026delays}. 

The structure of this article is as follows. In \secref{sec:theory} we review the virtual-cavity pulse formalism, the SLH rules used throughout the package and additional theoretical concepts of the package such as the correlation functions to determine the output modes, the interaction picture and pulse delay. In \secref{sec:implementation} we describe the implementation and the basic workflow, illustrated with a simple cavity-scattering tutorial. In \secref{sec:examples} we show several examples to emphasize the most important features of the package. Many further examples are available in the GitHub repository~\cite{qio_github} and the documentation~\cite{qio_documentation}.

\section{Theoretical background}

\label{sec:theory}
This section summarizes the theory behind the package. We begin with the input--output theory for quantum pulses, where the virtual-cavity construction represents wave packets as discrete oscillator modes~\cite{kiilerich2019inputoutput,kiilerich2020quantum}. Then the SLH formalism~\cite{combes2017slh} is introduced, which supplies algebraic rules for composing individual elements into cascaded networks. Next, we show how temporal output-mode bases follow from two-time correlation functions~\cite{kiilerich2019inputoutput,kiilerich2020quantum}. Finally, we cover two extensions: interaction-picture transformations for efficient simulations~\cite{christiansen2023interactions}, and virtual delay cavities for delayed propagation~\cite{christiansen2026delays}.

\subsection{Input--output theory with quantum pulses and virtual cavities}
The central goal of the input--output theory with quantum pulses~\cite{kiilerich2019inputoutput,kiilerich2020quantum} is to describe the dynamics of a photon pulse interacting with a localized quantum system~\cite{gheri1998wavepackets,gough2012filtering,baragiola2012nphoton}. In such a problem one is usually interested not only in the state of the system, but also in the temporal mode and quantum state of the incoming and outgoing radiation.

We consider a local quantum system with Hamiltonian $\hat{H}$ coupled to a traveling bosonic field $\hat b_{\rm in}(t)$ through a system operator $\hat c$ at rate $\kappa$. Under standard Markov and dispersionless-propagation assumptions~\cite{gardiner1985input,gardiner2004quantum}, the interaction can be written as
\begin{align}
\hat V_{\rm SB}(t)=i\sqrt{\kappa}\left(\hat c\,\hat b_{\rm in}^\dagger(t)-\hat c^\dagger\hat b_{\rm in}(t)\right),
\end{align}
and the outgoing field satisfies the input--output relation
\begin{align}
\hat{b}_{\rm out}(t)=S\,\hat{b}_{\rm in}(t)+\hat{L}(t),
\label{eq:io_relation}
\end{align}
where the coupling operator $\hat{L}$ is the field radiated by the system and the scattering matrix $S$ routes the incoming field directly into the outgoing one. For the single system considered here, $\hat{L}=\sqrt{\kappa}\,\hat{c}$ and there is no direct scattering, $S=1$, so that $\hat{b}_{\rm out}(t)=\hat{b}_{\rm in}(t)+\sqrt{\kappa}\,\hat{c}(t)$. Together with the system Hamiltonian $\hat{H}$, these form the triple $(S,\hat{L},\hat{H})$ that the SLH formalism of \secref{sec:slh} composes into networks.
The fields $\hat b_{\rm in}(t)$ and $\hat b_{\rm out}(t)$ carry one bosonic mode per time instant and thus span a continuum. A quantum pulse is a wave packet built from this continuum, its shape fixed by a normalized temporal mode $u(t)$ with $\int dt\,|u(t)|^2=1$.
This mode collects the continuum operators into a single bosonic mode, with creation operator
\begin{align}
\hat b_u^\dagger=\int dt\,u(t)\hat b^\dagger(t).
\end{align}
Even for a single incident mode, the interaction typically scatters the pulse into many outgoing modes, so the output field is spread over many modes. The aim is therefore to describe the full system--field dynamics in a tractable way, without explicitly discretizing the entire continuum.

The key idea is to replace the selected traveling input and output pulses by virtual cavities and to place the physical system between them, forming a \emph{cascade}, a directional chain in which the output field of one element drives the input of the next with no back-action from the elements downstream~\cite{gardiner1993driving,carmichael1993trajectory}. This network is sketched in \fref{fig:io_model}, with one input cavity and two output cavities surrounding the physical system. Because the coupling runs in one direction only, the ordering of the elements along the chain is physically meaningful.
Throughout this article, $u_i(t)$ denotes an input temporal mode and $v_i(t)$ an output temporal mode; the corresponding virtual-cavity annihilation operators are $\hat a_{u_i}$ and $\hat a_{v_i}$, respectively. 
Each virtual cavity is a discrete oscillator mode coupled to the continuum with a time-dependent, generally complex coupling determined by the desired temporal mode. For a normalized input mode $u(t)$, the coupling is~\cite{kiilerich2019inputoutput,kiilerich2020quantum}
\begin{align}
g_u(t)=\frac{u^*(t)}{\sqrt{1-\int_0^t dt'\,|u(t')|^2}}.
\label{eq:gu}
\end{align}
The coupling emits the initial intracavity state into the traveling mode $u(t)$. Conversely, for a chosen output mode $v(t)$, the coupling is~\cite{kiilerich2019inputoutput,kiilerich2020quantum}
\begin{align}
g_v(t)=-\frac{v^*(t)}{\sqrt{\int_0^t dt'\,|v(t')|^2}}.
\label{eq:gv}
\end{align}
This coupling absorbs the mode into a virtual output cavity.
These expressions make it possible to treat selected input and output pulses as single oscillator modes in a cascaded network.
The couplings $g_u(t)$ and $g_v(t)$ can be intuitively understood by noting that the squared denominators in \eqref{eq:gu} and \eqref{eq:gv} give the fractions of the pulse residing inside the emitting and absorbing virtual cavities, respectively~\cite{christiansen2026delays}. The numerators specify the desired temporal envelopes, while the denominators account for the instantaneous intracavity photon populations during emission and absorption. 

For multiple input or output modes, the presence of the other virtual cavities distorts the mode seen by a given
cavity, which needs to be taken into account~\cite{kiilerich2019inputoutput,kiilerich2020quantum}. One therefore first constructs effective modes $u_i^{\rm eff}(t)$ and $v_i^{\rm eff}(t)$. For input modes, the recursive construction for $j<i$ is
\begin{align}
\dot\alpha_{u_i}^{(j)}(t)&=-g_{u_j}(t)u_i^{(j-1)}(t)+\frac{1}{2}|g_{u_j}(t)|^2\alpha_{u_i}^{(j)}(t),\\
u_i^{(j)}(t)&=u_i^{(j-1)}(t)-g_{u_j}^*(t)\alpha_{u_i}^{(j)}(t).
\end{align}
For output modes, the corresponding recursion is
\begin{align}
\dot\alpha_{v_i}^{(j)}(t)&=-g_{v_j}(t)v_i^{(j-1)}(t)-\frac{1}{2}|g_{v_j}(t)|^2\alpha_{v_i}^{(j)}(t),\\
v_i^{(j)}(t)&=v_i^{(j-1)}(t)+g_{v_j}^*(t)\alpha_{v_i}^{(j)}(t).
\end{align}
With $u_i^{(0)}(t)=u_i(t)$, $v_i^{(0)}(t)=v_i(t)$, and $\alpha_{u_i}^{(j)}(0)=\alpha_{v_i}^{(j)}(0)=0$, the effective modes are $u_i^{\rm eff}(t)=u_i^{(i-1)}(t)$ and $v_i^{\rm eff}(t)=v_i^{(i-1)}(t)$. The corresponding couplings follow from the single-mode expressions,
\begin{align}
g_{u_i}^{\rm eff}(t)&=\frac{\left(u_i^{\rm eff}(t)\right)^*}{\sqrt{1-\int_0^t dt'\,|u_i^{\rm eff}(t')|^2}},\\
g_{v_i}^{\rm eff}(t)&=-\frac{\left(v_i^{\rm eff}(t)\right)^*}{\sqrt{\int_0^t dt'\,|v_i^{\rm eff}(t')|^2}}.
\end{align}
This recursive construction gives the effective temporal modes that determine the virtual-cavity couplings in a multimode description. We note that the dynamics of the effective couplings can be precalculated independently of the system dynamics.

\subsection{SLH networks for cascaded pulses}
\label{sec:slh}

\begin{figure}
    \centering 
    \includegraphics[width=\columnwidth]{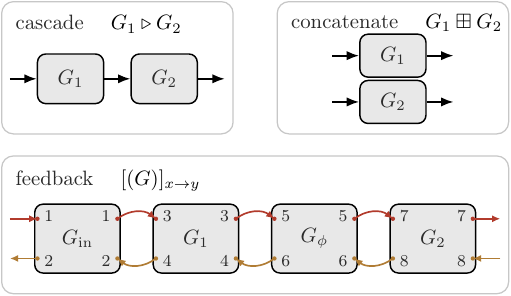}
    \caption{
      SLH composition rules. \emph{Cascade} ($G_1\triangleright G_2$) feeds output into input; \emph{concatenate} ($G_1\boxplus G_2$) stacks components without interaction; \emph{feedback reduction} ($[(G)]_{x\to y}$) closes a loop from output $x$ to input $y$. The feedback panel is the bidirectional waveguide of \secref{sec:feedback} (\coderef{code:feedback}): input coupler $G_{\rm in}$, emitters $G_1$, $G_2$, and propagation phase $G_\phi$. The numbers label the ports of the concatenated network $G_{\rm in}\boxplus G_1\boxplus G_\phi\boxplus G_2$, each marking both the input and the output of that port. Red and orange lines mark the right- and left-moving channels.
    }
    \label{fig:SLH_rules}
\end{figure}

The virtual-cavity construction of the previous subsection turns the pulse problem into a small collection of open components: input cavities that emit the incident modes, the physical system, and one output cavity per captured mode. These components form a cascade, but the dynamics follow from the master equation of the full chain, not of each part on its own. Deriving it by hand is tedious and must be redone for every new arrangement of components. The SLH formalism~\cite{gough2009hamiltonian,combes2017slh} removes this bookkeeping: each open component is encoded in a triple $G=(S,\hat{L},\hat{H})$, and a few algebraic rules combine these building blocks into a single triple for the whole network, which maps directly onto a Lindblad master equation.

The input--output relation \eqref{eq:io_relation} already introduced these objects for a single system, with coupling operator $\hat{L}=\sqrt{\kappa}\,\hat{c}$ and scattering matrix $S=1$. An SLH component generalizes them to several ports, or field channels: the coupling vector $\hat{L}=(\hat{L}_1,\dots,\hat{L}_n)^\top$ carries one jump operator $\hat{L}_j$ per port, the scattering matrix $S$ becomes an $n\times n$ unitary that routes the incoming fields directly to the outgoing ports, and $\hat{H}$ is its internal Hamiltonian. Given a triple, the state $\rho$ evolves under the Lindblad master equation
\begin{align}
\dot{\rho}=-i\,[\hat{H},\rho]+\sum_j\left(\hat{L}_j\rho\,\hat{L}_j^\dagger-\tfrac{1}{2}\big\{\hat{L}_j^\dagger\hat{L}_j,\rho\big\}\right),
\label{eq:slh_me}
\end{align}
set entirely by $\hat{H}$ and $\hat{L}$. The scattering matrix $S$ does not enter here; it fixes how the fields scatter between ports through the input--output relation \eqref{eq:io_relation}, and how components combine into a network. A component that only emits or absorbs keeps $S=1$: a one-sided cavity with decay rate $\kappa$, detuning $\Delta$, and mode operator $\hat{c}$ has the SLH triple $(1,\sqrt{\kappa}\,\hat{c},-\Delta\hat{c}^\dagger\hat{c})$, while an input cavity emitting the mode $u(t)$ is $(1,g_u^*(t)\hat{a}_u,0)$ with its coupling set by the time-dependent $g_u(t)$ of \eqref{eq:gu}. In \appref{sec:SLH_elements}, we provide a list of common SLH elements, see also the appendix of Ref.~\cite{combes2017slh}. Throughout, we denote the cavity decay rate by $\kappa$ and that of an emitter by $\gamma$.

Three composition rules, illustrated in \fref{fig:SLH_rules}, assemble these components into a network. The cascade product $G_1\triangleright G_2$ connects the output of $G_1$ to the input of $G_2$ and gives the combined triple
\begin{align}
G_1\triangleright G_2=\Big( &S_2S_1,\,\hat{L}_2+S_2\hat{L}_1,\,  \notag \\
&\hat{H}_1+\hat{H}_2+\frac{1}{2i}\big(\hat{L}_2^\dagger S_2\hat{L}_1-\hat{L}_1^\dagger S_2^\dagger \hat{L}_2\big) \Big).
\label{eq:cascade}
\end{align}
Note that we use the convention of cascading elements from left to right, $G_1\triangleright G_2$, which differs from the convention in Ref.~\cite{combes2017slh}, where $G_2 \triangleleft G_1$ is used.
Next, the concatenation product $G_1\boxplus G_2$ stacks channels side by side without interaction,
\begin{align}
G_1\boxplus G_2=\left(\begin{bmatrix}S_1&0\\0&S_2\end{bmatrix},\,\begin{bmatrix}\hat{L}_1\\\hat{L}_2\end{bmatrix},\,\hat{H}_1+\hat{H}_2\right).
\label{eq:concat}
\end{align}
A third rule, feedback reduction, connects output port $x$ back into input port $y$ of an $n$-port component. The reduced $(n-1)$-port model $[(G)]_{x\to y}=(\tilde S,\hat{\tilde L},\hat{\tilde H})$ is
\begin{align}
\tilde S&=S_{\bar x,\bar y}+S_{\bar x,y}(1-S_{x,y})^{-1}S_{x,\bar y},
\label{eq:fbS}\\
\hat{\tilde L}&=\hat{L}_{\bar x}+S_{\bar x,y}(1-S_{x,y})^{-1}\hat{L}_x,
\label{eq:fbL}\\
\hat{\tilde H}&=\hat{H} - \frac{i}{2}\left[\left(\sum_{j=1}^n\hat{L}_j^\dagger S_{j,y}\right)(1-S_{x,y})^{-1}\hat{L}_x-\mathrm{h.c.}\right].
\label{eq:fbH}
\end{align}
Here $\bar x$ and $\bar y$ denote the remaining output and input ports after removing $x$ and $y$. This rule closes internal loops in an SLH network and is used whenever it contains a direct coherent feedback path. The feedback reduction in combination with the concatenation rule is sufficient to describe any SLH network. However, in many cases the cascade product is simpler and more efficient to implement.

By modeling input and output pulses as virtual cavities and cascading them with the physical system, we obtain an effective SLH triple for the full pulse-interaction problem. 
Its Hamiltonian $\hat{H}$ and coupling operators $\hat{L}$ then determine the dynamics of the joint state of the system and the auxiliary pulse modes through \eqref{eq:slh_me}, which can be solved with standard Lindblad solvers~\cite{Johansson2012,Johansson2013,Kraemer2018,QuantumToolbox2025}.

\subsection{Output modes and the correlation function}
Solving \eqref{eq:slh_me} yields the joint state of the system and the pulse modes, but in many cases the quantity of interest is the emitted field. It is generally spread across a continuum of temporal modes, so before it can be captured by a finite set of output cavities we must identify which ones it occupies. These modes are singled out by the first-order correlation function
\begin{align}
g^{(1)}(t_1,t_2)=\langle \hat{L}_s^\dagger(t_1)\hat{L}_s(t_2)\rangle,
\label{eq:g1}
\end{align}
obtained from the master-equation solution through the quantum regression theorem. Here $\hat{L}_s$ is the coupling operator of the source, obtained by cascading the input cavity with the system and evaluated without the output cavities. For a one-sided cavity with annihilation operator $\hat{c}$ and decay rate $\kappa$ driven by a pulse $u(t)$, this gives $\hat{L}_s(t)=g_u^*(t)\hat{a}_u+\sqrt{\kappa}\,\hat{c}$: the system term $\sqrt{\kappa}\,\hat{c}$ plus the input-cavity term $g_u^*(t)\hat{a}_u$ added by the cascade.

Diagonalizing the correlation function,
\begin{align}
g^{(1)}(t_1,t_2)=\sum_i n_i\, v_i^*(t_1)v_i(t_2),
\label{eq:g1_decomp}
\end{align}
yields an orthonormal set of temporal modes $v_i(t)$ ranked by their mean occupation $n_i$. 

For many applications, the few most populated modes may contain the quantum states of our primary interest, and are hence the modes the output cavities are set to absorb, with couplings built from the $v_i(t)$ through \eqref{eq:gv}. This gives the working recipe used throughout the examples: a first propagation without output cavities provides $g^{(1)}$ and its modes, and a second propagation, with cavities tuned to the leading $v_i(t)$, turns the output field into explicit quantum degrees of freedom (\secref{sec:implementation}). 

\subsection{Interaction picture for traveling pulses}
\label{sec:interaction_picture}
For pulse-scattering problems, especially with large photon numbers, it is often advantageous to transform to an interaction picture that treats the pure mode-transfer dynamics analytically between the input and output modes~\cite{christiansen2023interactions}. To this end, one splits the Hamiltonian into the system part and a coupling part that only exchanges excitations between upstream and downstream virtual modes. For the two-mode case this mode-transfer Hamiltonian is
\begin{align}
\hat{H}_{uv}(t)=\frac{i}{2}\left[g_u(t)g_v^*(t)\hat{a}_u^\dagger\hat{a}_v-g_v(t)g_u^*(t)\hat{a}_v^\dagger\hat{a}_u\right].
\end{align}
The transformation acts only on the field-mode operators,
\begin{align}
\begin{pmatrix}\hat{a}_u(t)\\ \hat{a}_v(t)\end{pmatrix}=M(t)\begin{pmatrix}\hat{a}_u(0)\\ \hat{a}_v(0)\end{pmatrix},
\end{align}
and the coefficient matrix solves
\begin{align}
\frac{d}{dt}M(t)&=A(t)M(t),\\
A(t)&=\frac{1}{2}\begin{bmatrix}0&g_u(t)g_v^*(t)\\-g_u^*(t)g_v(t)&0\end{bmatrix},
\end{align}
with $M(0)=I$. The generator $A(t)$ is the one induced by the mode-transfer Hamiltonian $\hat{H}_{uv}$. In this picture, $\hat{H}_{uv}$ is subtracted before the transformed mode operators are substituted into the remaining Hamiltonian. The same substitution transforms the Lindblad coupling operators. When the scatterer exchanges only a limited number of excitations with the transformed modes, their occupations remain within narrow Fock-state bands, allowing smaller Hilbert-space truncations.

For $N$ cascaded modes the interaction-picture transformation is generated by the matrix
\begin{align}
A(t)=\frac{1}{2}\begin{bmatrix}
0&g_1g_2^*&\cdots&g_1g_N^*\\
-g_1^*g_2&0&\ddots&\vdots\\
\vdots&\ddots&0&g_{N-1}g_N^*\\
-g_1^*g_N&\cdots&-g_{N-1}^*g_N&0
\end{bmatrix},
\end{align}
which corresponds to the interaction Hamiltonian obtained from $G_1\triangleright\cdots\triangleright G_N$, where $G_i=(1,g_i^* \hat{a}_i,0)$.

\subsection{Pulse delay}
Propagation delays must be accounted for when pulses traverse paths of different lengths, as in interferometric networks, where only the relative delay between paths is physically relevant. Resolving such a delay directly would require discretizing the field along its propagation direction, that is, the full continuum. Instead, a single virtual delay cavity captures the pulse and re-emits it later, adding the delay with one extra mode~\cite{christiansen2026delays}.

When the delay is long, the capture and emission windows do not overlap: the delay cavity first absorbs the incoming pulse $v(t)$ through an absorbing coupling as in \eqref{eq:gv}, then releases it as $u(t)$ through an emitting coupling as in \eqref{eq:gu}. When the delay is short, the two windows overlap, and the delay cavity absorbs and emits at the same time. We therefore model it as a two-sided cavity: the pulse $v(t)$ enters through an input mirror, while the delayed pulse $u(t)$ leaves through a separate output mirror. The two mirrors have the time-dependent couplings
\begin{align}
\tilde g_{{\rm out},u,v}(t)&=\frac{u^*(t)}{\sqrt{\int_0^t dt'\,|v(t')|^2-\int_0^t dt'\,|u(t')|^2}},
\label{eq:gout_delay}\\
\tilde g_{{\rm in},v,u}(t)&=-\frac{v^*(t)}{\sqrt{\int_0^t dt'\,|v(t')|^2-\int_0^t dt'\,|u(t')|^2}},
\label{eq:gin_delay}
\end{align}
for the output and input mirrors, respectively. The shared denominator, the captured norm minus the released norm, is the fraction of the pulse stored in the cavity at time $t$. Interferometric settings then follow by placing one delay cavity per path and tuning the relative delays. 

\section{Framework and implementation}
\label{sec:implementation}

\begin{figure}
    \centering 
    \includegraphics[width=\columnwidth]{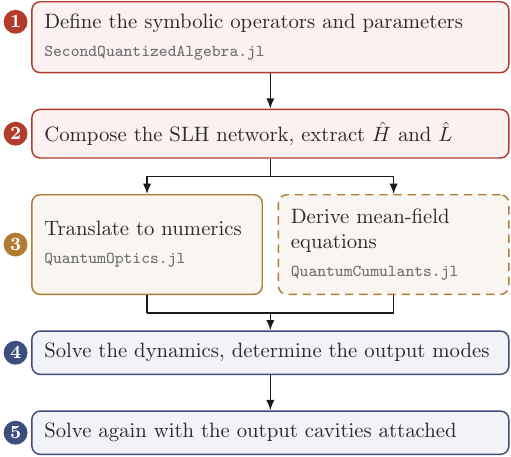}
    \caption{Workflow sketch for QuantumInputOutput.jl. The symbolic operators are defined to build an SLH network before translating them to numeric operators or deriving higher-order mean-field equations. The resulting Hamiltonian and Lindblad operators are then used to simulate the dynamics and determine the output modes. Finally, the full dynamics with output cavities are solved. The translation can be avoided by directly defining the numeric operators.}
    \label{fig:workflow}
\end{figure}

The central object is the \code{SLH} struct, representing the triple $(S,\hat{L},\hat{H})$. It supports the network operations \code{cascade} (\codesymbol{$\triangleright$}) [\eqref{eq:cascade}], \code{concatenate} (\codesymbol{$\boxplus$}) [\eqref{eq:concat}], and \code{feedback} [\eqsref{eq:fbS}--(\ref{eq:fbH})], introduced in \secref{sec:slh} and illustrated in \fref{fig:SLH_rules}. The entries of an \code{SLH} object may be symbolic expressions, numerical operators, or operator-valued functions, so the same composition rules are used in both workflows. 
From a composite SLH object, the effective operators are extracted with the functions \code{hamiltonian}, \code{jump\_operator}, and \code{scattering}.

In the symbolic workflow, the SLH rules produce symbolic Hamiltonian and Lindblad coupling operators. These expressions can then be represented numerically for a full master-equation simulation with \texttt{QuantumOptics.jl}~\cite{Kraemer2018}, or used directly to derive higher-order mean-field equations for operator averages with \texttt{QuantumCumulants.jl}~\cite{Plankensteiner2022quantumcumulantsjl}. The symbolic expressions can also be inspected, simplified, or further manipulated before a numerical representation is chosen. Alternatively, numerical functions and operators can be provided directly, bypassing the symbolic derivation, as demonstrated for the $\Lambda$-emitter in \coderef{code:lambda_numeric} of \secref{sec:lambda_modes}.

The five stages of this workflow are summarized in \fref{fig:workflow}. Stages 1 and 2 define the symbolic operators and compose the SLH network. Stage 3 maps the resulting operators to a numerical representation or derives cumulant equations. Stage 4 solves the dynamics and identifies the relevant output modes. Stage 5 repeats the propagation with virtual cavities that capture these modes. The symbolic translation in stage 3 can instead be bypassed by defining the SLH components with numerical operators from the outset.

In the following, we illustrate the main concept in more detail using a simple cavity-scattering example. A virtual cavity emits a single input pulse, which is scattered by a one-sided cavity and recaptured by a virtual output cavity \cite{kiilerich2019inputoutput}. First, we solve the dynamics of the input pulse and the system to extract the dominant outgoing temporal mode from the two-time correlation function. We then represent this mode explicitly by an output cavity with coupling $g_v(t)$. This two-stage procedure is typical of the virtual-cavity formalism for analyzing output modes.

For this tutorial, we use the symbolic route provided by \texttt{SecondQuantizedAlgebra.jl}, which defines noncommuting operators and symbolic parameters and manipulates the resulting expressions. In \coderef{code:symbolic} we construct three bosonic modes on separate \code{FockSpace} factors: the virtual input-cavity mode $\hat{a}_u$, the physical system-cavity mode $\hat{c}$, and the virtual output-cavity mode $\hat{a}_v$. We encode these components as SLH elements and \code{cascade} them to derive the symbolic Hamiltonian and Lindblad coupling operator of the complete pulse-scattering model.
The same procedure generalizes directly to larger networks: one combines local elements with cascade and concatenation rules, or applies feedback reduction when internal loops are present.

\begin{lstlisting}[caption={Defining the symbolic operators and parameters for the SLH construction of a single input mode, one cavity, and one output mode.},label={code:symbolic}]
using QuantumInputOutput
using SecondQuantizedAlgebra

# Hilbert space
hᵤ = FockSpace(:u)
hₛ = FockSpace(:s)
hᵥ = FockSpace(:v)
h = hᵤ ⊗ hₛ ⊗ hᵥ

# symbolic operators
aᵤ = Destroy(h, :aᵤ, 1)
c = Destroy(h, :c, 2)
aᵥ = Destroy(h, :aᵥ, 3)

# symbolic parameters
@variables gᵤ::Number gᵥ::Number Δ::Real κ::Real

# SLH elements
Gᵤ = SLH(1, gᵤ'*aᵤ, 0)
Gₛ = SLH(1, √κ*c, -Δ*c'*c)
Gᵥ = SLH(1, gᵥ'*aᵥ, 0)

# cascade network
G = Gᵤ ▷ Gₛ ▷ Gᵥ
H = hamiltonian(G)
L = jump_operator(G)[1]
\end{lstlisting}
The above code sample creates the symbolic expressions 
\begin{align}
\hat{H} &= -\Delta \hat{c}^\dagger\hat{c}
 + \frac{i}{2}\Big[
    \sqrt{\kappa}g_u\hat{a}_u^\dagger\hat{c}
    + g_u g_v^*\hat{a}_u^\dagger\hat{a}_v
 \notag\\
 &\quad
    + \sqrt{\kappa}g_v^*\hat{c}^\dagger\hat{a}_v
  \Big]
 + \mathrm{h.c.},
 \nonumber\\
\hat{L} &= g_u^*\hat{a}_u+\sqrt{\kappa}\hat{c}
     +g_v^*\hat{a}_v.
\end{align}
The three interaction terms in $\hat H$ are generated by the successive cascade products and describe the directional couplings from the input cavity to the system, from the input cavity to the output cavity, and from the system to the output cavity. The coupling operator $\hat L$ collects the amplitudes with which the three components couple to the shared traveling field. Because the network has one field channel, it produces one Lindblad coupling operator.

We now use \code{to\_numeric} to represent the symbolic Hamiltonian and coupling operator on a \texttt{QuantumOptics.jl} basis. The \code{parameter} dictionary supplies constant values, whereas \code{time\_parameter} supplies functions of time. Since the input coupling $g_u(t)$ is time dependent, the translated Hamiltonian and coupling operator are callable objects that are evaluated at the current time by the solver callback.

The network contains a single excitation, so a \code{FockBasis} with a cutoff of one is sufficient for each of the three modes. For the first propagation, we place $g_v=0$ in the constant-parameter dictionary so that the output cavity remains in the numerical Hilbert space but is dynamically decoupled. The function \code{coupling\_input} constructs $g_u(t)$ from the Gaussian input mode according to \eqref{eq:gu}.

\begin{lstlisting}[caption={Translation to numerical operators with a time-dependent Gaussian input coupling.},label={code:translation}]
using QuantumOptics

# numeric basis
bᵤ = FockBasis(1)
bₛ = FockBasis(1)
bᵥ = FockBasis(1)
b = bᵤ ⊗ bₛ ⊗ bᵥ

# input coupling
T = [0.0:0.002:12.0;]
u(t) = 1/π^(1/4)*exp(-0.5*(t - 4)^2)
gᵤ_qo = coupling_input(u, T)

# constant and time-dependent parameters
parameter = Dict(κ => 1.0, Δ => 0.0, gᵥ => 0.0)
time_parameter = Dict(gᵤ => gᵤ_qo)

# numeric time-dependent H and jump operator
H_qo = to_numeric(H, b; parameter, time_parameter)
L_qo = to_numeric(L, b; parameter, time_parameter)
\end{lstlisting}

The time-dependent operators returned by \code{to\_numeric} can now be used with the \texttt{QuantumOptics.jl} function \code{timeevolution.master\_dynamic}. At each integration step, the solver calls $f_{\rm io}(t,\rho)$, which evaluates the translated Hamiltonian and coupling operators and returns the Hamiltonian, the vector of Lindblad coupling operators, and their adjoints. In \coderef{code:evolution} we use this callback to propagate the single-photon pulse.

\begin{lstlisting}[caption={Solving the time-dependent master equation for the cavity-scattering example.},label={code:evolution}]
# input-output function
function f_io(t, ρ)
  Ht = H_qo(t)
  Lt = [L_qo(t)]
  return Ht, Lt, dagger.(Lt)
end

# initial state
ψ₀ =
  fockstate(bᵤ, 1) ⊗ fockstate(bₛ, 0) ⊗ fockstate(bᵥ, 0)

# time evolution
t, ρt = timeevolution.master_dynamic(T, ψ₀, f_io)
\end{lstlisting}

\begin{figure}[!t] 
\centering
\includegraphics[width=0.99\columnwidth]{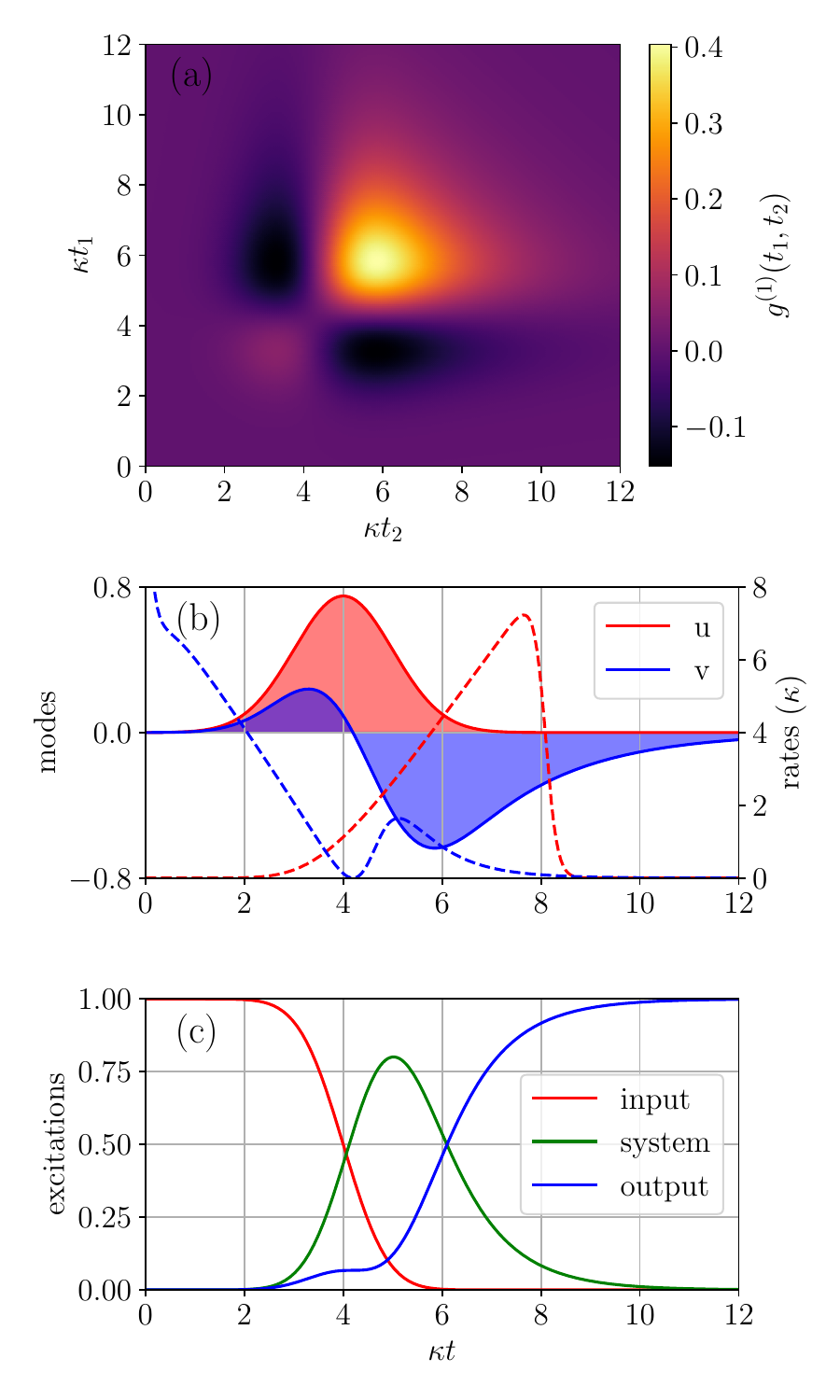}
\caption{Empty cavity scattering. Panel (a) shows the two-time correlation function of the emitted field. Its dominant eigenvector defines the temporal output mode used in the second propagation step. Panel (b) shows the temporal input and output modes with their virtual-cavity couplings (dashed) and (c) depicts the dynamics of the occupations of the input cavity, system cavity, and captured output mode.}
\label{fig:tutorial}
\end{figure}

The source coupling operator $\hat{L}_s(t)=g_u^*(t)\hat{a}_u+\sqrt{\kappa}\,\hat{c}$ of \eqref{eq:g1} describes the field emitted by the input and system cavities. The function \code{correlation\_matrix} uses this operator and the trajectory $\rho(t)$ to construct the first-order correlation matrix shown in \fref{fig:tutorial}(a).

Diagonalizing the time-grid-discretized matrix $g^{(1)}(t_1,t_2)$ yields the orthogonal output modes and their occupations. The dominant eigenvector defines the mode function $v(t)$, from which \code{coupling\_output} constructs the absorption coupling $g_v(t)$ according to \eqref{eq:gv}, as shown in \coderef{code:correlation}. The eigenvalues and corresponding eigenvectors are sorted in ascending order.

\begin{lstlisting}[caption={Extracting the dominant temporal output mode from the two-time correlation matrix.},label={code:correlation}]
using LinearAlgebra

# two-time correlation function
Lₛ(t) = dense(static_operator(L_qo(t)))
g¹ = correlation_matrix(T, ρt, f_io, Lₛ)

# diagonalize the correlation function
F = eigen(g¹)
ΔT = T[2] - T[1]

# eigenvectors (temporal modes)
v_mode = F.vectors[:, end] / √ΔT
# corresponding mean photon occupation
n_mode = real(F.values[end])*ΔT
\end{lstlisting}

In the simple linear scattering of an empty cavity, all population ends up in a single dominant mode. Retranslating the same symbolic Hamiltonian and Lindblad term with both $g_u(t)$ and $g_v(t)$ then produces the full effective model in which the output mode is an explicit quantum degree of freedom. This second propagation step is shown in \coderef{code:output_mode}.

\begin{lstlisting}[caption={Time evolution with the output mode.},label={code:output_mode}]
# coupling of the output mode
gᵥ_qo = coupling_output(v_mode, T)

# translate again, with input and output coupling
parameter = Dict(κ => 1.0, Δ => 0.0)
time_parameter = Dict(gᵤ => gᵤ_qo, gᵥ => gᵥ_qo)
H_qo = to_numeric(H, b; parameter, time_parameter)
L_qo = to_numeric(L, b; parameter, time_parameter)

# f_io reads the reassigned operators
t, ρt = timeevolution.master_dynamic(T, ψ₀, f_io)

# numeric operators and expectation values
aᵤ_qo = to_numeric(aᵤ, b)
c_qo = to_numeric(c, b)
aᵥ_qo = to_numeric(aᵥ, b)

nᵤ = real.(expect(aᵤ_qo'*aᵤ_qo, ρt))
nₛ = real.(expect(c_qo'*c_qo, ρt))
nᵥ = real.(expect(aᵥ_qo'*aᵥ_qo, ρt))
\end{lstlisting}

In \fref{fig:tutorial}(b) we plot the temporal input and output modes together with their corresponding couplings.
Figure \ref{fig:tutorial}(c) shows their excitations calculated in the preceding code sample, see also Ref.~\cite{kiilerich2019inputoutput}.
We can see that the virtual output cavity captures the dominant temporal mode identified from the correlation matrix.

\section{Examples} \label{sec:examples}
The above tutorial demonstrated the main workflow of the package.
In the following, we show several examples
that emphasize additional features of the package. In
particular, \secref{sec:lambda_modes} covers multilevel systems, multiple captured output modes, user-defined operator translations via the \code{operators} keyword, and a fully numerical workflow without symbolic derivation. Section \ref{sec:feedback} demonstrates bidirectional channels, the \code{concatenate} and \code{feedback} rules, and coherent driving with a time-dependent pulse envelope. Finally, \secref{sec:interaction_picture_example} shows the interaction-picture treatment for highly occupied quantum pulses and a cumulant-expansion approach. 

\subsection{\texorpdfstring{Photon number and emitter--mode entanglement with a $\Lambda$ emitter}{Photon number and emitter--mode entanglement with a Lambda emitter}}
\label{sec:lambda_modes}

As a first example, we consider the decay of a three-level $\Lambda$ emitter through a cavity into two dominant temporal output modes~\cite{kiilerich2020quantum}. The emitted field is multimode. We determine the dominant modes from the first-order correlation function and retain the two most populated ones as explicit output cavities. This gives direct access to correlations and entanglement between the emitter and the selected propagating modes. 

The symbolic model consists of one cavity mode, one three-level atom, and two cascaded virtual output cavities. There is no input cavity, because the emitter begins in $\ket{e}$ and the physical cavity in vacuum.
Levels 1, 2, and 3 represent $\ket{g_1}$, $\ket{g_2}$, and $\ket{e}$, respectively.
Here the transitions $\ket{g_1}\leftrightarrow\ket{e}$ and $\ket{g_2}\leftrightarrow\ket{e}$ are both coupled to the cavity with strength $g$, while the second ground state is shifted by $\omega_{12}$. The two virtual cavities later absorb the two dominant temporal output modes. All components couple to the same traveling field channel.
In \coderef{code:lambda_symbolic} we define the symbolic operators and parameters and derive the Hamiltonian and Lindblad operator.

\begin{lstlisting}[caption={Symbolic SLH model for a three-level emitter coupled to a cavity with two output modes.},label={code:lambda_symbolic}]
using QuantumInputOutput
using SecondQuantizedAlgebra

# symbolic Hilbert space
hc = FockSpace(:c)
hₐ = NLevelSpace(:atom, 3)
hᵥ₁ = FockSpace(:v1)
hᵥ₂ = FockSpace(:v2)
h = hc ⊗ hₐ ⊗ hᵥ₁ ⊗ hᵥ₂

# symbolic operators
c = Destroy(h, :c, 1)
σ(i, j) = Transition(h, :σ, i, j, 2)
aᵥ₁ = Destroy(h, :aᵥ₁, 3)
aᵥ₂ = Destroy(h, :aᵥ₂, 4)

# symbolic parameters
@variables κ::Real g::Real ω₁₂::Real
@variables gᵥ₁::Number gᵥ₂::Number

# Hamiltonian and SLH elements
Hₛ =
  g*(c'*σ(1, 3) + c*σ(3, 1) + c'*σ(2, 3) + c*σ(3, 2)) +
  ω₁₂*σ(2, 2)
Gₛ = SLH(1, √κ*c, Hₛ)
Gᵥ₁ = SLH(1, gᵥ₁'*aᵥ₁, 0)
Gᵥ₂ = SLH(1, gᵥ₂'*aᵥ₂, 0)

# SLH network
G = cascade(Gₛ, Gᵥ₁, Gᵥ₂)
H = hamiltonian(G)
L = jump_operator(G)[1]
\end{lstlisting}

We first determine the two dominant output modes from the emitter--cavity trajectory. Since the output cavities are downstream, they cannot affect the upstream emitter--cavity dynamics. We therefore omit them from the first propagation and retain only the physical cavity and the emitter. 
The \code{operators} dictionary maps the symbolic system operators onto this reduced numerical basis. The resulting Hamiltonian and coupling operator are time independent and are propagated with \code{timeevolution.master}. Diagonalizing the first-order correlation matrix then yields the temporal modes and their mean occupations. Since only the two most populated modes are needed, we compute only the five dominant eigenvectors, rather than the full spectrum.

\begin{lstlisting}[caption={System time evolution with a reduced numerical Hilbert space using the operators dictionary to determine the output modes.},label={code:translation_2modes}]
using QuantumOptics

# numerical parameters
κ_qo = 1.0
g_qo = 0.1*κ_qo
ω₁₂_qo = 0.5*κ_qo
T = [0:0.005:1;] * 100/κ_qo
ΔT = T[2] - T[1]

# system basis: cavity and three-level atom
bc = FockBasis(1)
bₐ = NLevelBasis(3)
bₛ = bc ⊗ bₐ

# full basis
bᵥ₁ = FockBasis(1)
bᵥ₂ = FockBasis(1)
b = bₛ ⊗ bᵥ₁ ⊗ bᵥ₂

# operators on the system basis
cₛ = destroy(bc) ⊗ one(bₐ)
σₛ(i, j) = one(bc) ⊗ transition(bₐ, i, j)

# parameter and operator dictionaries
parameter = Dict(κ => κ_qo, g => g_qo,
  ω₁₂ => ω₁₂_qo, gᵥ₁ => 0.0, gᵥ₂ => 0.0)
operators = Dict(c => cₛ, σ(1, 3) => σₛ(1, 3),
  σ(2, 3) => σₛ(2, 3), σ(2, 2) => σₛ(2, 2))

# numeric (static) operators on the system basis
Hₛ_qo = to_numeric(Hₛ, bₛ; parameter, operators)
Lₛ_qo = to_numeric(√κ*c, bₛ; parameter, operators)

# system time evolution
ψ₀ₛ = fockstate(bc, 0) ⊗ nlevelstate(bₐ, 3)
t₁, ρt₁ = timeevolution.master(T, ψ₀ₛ, Hₛ_qo, [Lₛ_qo])

# correlation function and output modes
g¹ = correlation_matrix(T, ρt₁, Hₛ_qo, [Lₛ_qo], Lₛ_qo)
n = size(g¹, 1)
F = eigen(g¹, (n - 4):n)
n_avg = real.(F.values) * ΔT
v₁_mode = F.vectors[:, end] / √ΔT
v₂_mode = F.vectors[:, end-1] / √ΔT
n₁ = n_avg[end]
n₂ = n_avg[end-1]
\end{lstlisting}

For serial capture, the first absorbing cavity modifies the field seen by the second. The function \code{effective\_output\_mode} therefore transforms $v_2(t)$ into an effective mode that compensates for this distortion.
We then complete the example in the same way as in the tutorial: retranslate the full symbolic Hamiltonian and Lindblad operator with both output couplings included, and solve the second propagation step with the two explicit output modes:
\begin{lstlisting}[caption={Time evolution including the output cavities with the standard symbolic-to-numerical translation.},label={code:lambda_translate_full}]
# output coupling for two modes
gᵥ₁_qo = coupling_output(v₁_mode, T)
v₂_eff = effective_output_mode([v₁_mode, v₂_mode], T, 2)
gᵥ₂_qo = coupling_output(v₂_eff, T)

parameter = Dict(κ => κ_qo, g => g_qo, ω₁₂ => ω₁₂_qo)
time_parameter = Dict(gᵥ₁ => gᵥ₁_qo, gᵥ₂ => gᵥ₂_qo)

# full Hamiltonian and jump operator, now on b
H_qo = to_numeric(H, b; parameter, time_parameter)
L_qo = to_numeric(L, b; parameter, time_parameter)

function f_io(t, ρ)
  Lt = [L_qo(t)]
  return H_qo(t), Lt, dagger.(Lt)
end

# initial state and time evolution
ψ₀ = ψ₀ₛ ⊗ fockstate(bᵥ₁, 0) ⊗ fockstate(bᵥ₂, 0)
t₂, ρt₂ = timeevolution.master_dynamic(T, ψ₀, f_io)

# expectation values
c_qo = to_numeric(c, b)
σ_qo(i, j) = to_numeric(σ(i, j), b)
aᵥ₁_qo = to_numeric(aᵥ₁, b)
aᵥ₂_qo = to_numeric(aᵥ₂, b)

Pₑ = real.(expect(σ_qo(3, 3), ρt₂))
nc = real.(expect(c_qo'*c_qo, ρt₂))
n₁t = real.(expect(aᵥ₁_qo'*aᵥ₁_qo, ρt₂))
n₂t = real.(expect(aᵥ₂_qo'*aᵥ₂_qo, ρt₂))

# emitter--mode entanglement at the final time
ρ_atom_modes = ptrace(ρt₂[end], 1)
N_atom_modes = negativity(ρ_atom_modes, 1)
\end{lstlisting}

\begin{figure*}
\centering
\includegraphics[width=0.99\textwidth]{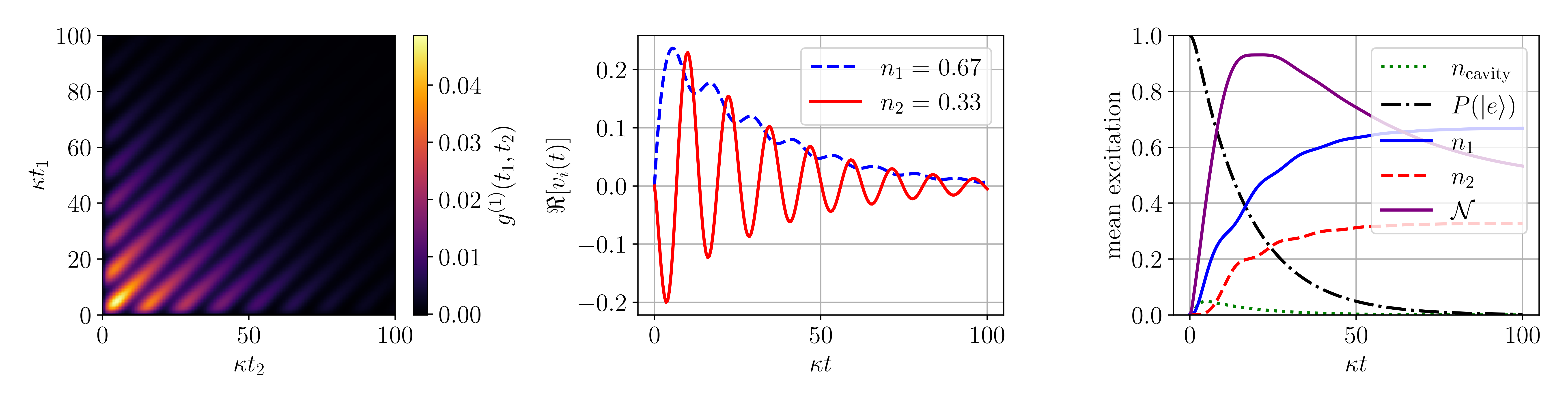}
\caption{Photon number and emitter--mode entanglement for a three-level $\Lambda$ emitter. The left panel shows the first-order correlation matrix of the emitted field obtained from the reduced cavity--atom simulation. The middle panel shows the real parts of the two dominant output modes extracted from the two eigenvectors with the largest eigenvalues of the correlation matrix. These modes are complex, and their eigenvectors are defined only up to a global phase, so each is rotated by the phase that maximizes its real part. The right panel shows the time evolution of the cavity population, the excited-state population, the occupations of the two captured output modes, and the emitter--mode negativity $\mathcal{N}$ during the second propagation step.}
\label{fig:lambda_modes}
\end{figure*}

The resulting observables are shown in \fref{fig:lambda_modes}, see also Ref.~\cite{kiilerich2020quantum}. The first panel shows the two-time correlation function of the emitted field. The second panel shows the temporal envelopes of the two dominant modes,
and the third panel demonstrates how the initial atomic excitation is transferred into the cavity and finally into the two captured output modes. The agreement $n_1\approx n_{1t}(T)$ and $n_2\approx n_{2t}(T)$ confirms that both modes identified from the correlation matrix are captured by the virtual cavities.
The final density operator retains the joint state of the emitter and the captured temporal modes, from which photon-number correlations and emitter--mode entanglement can be evaluated. Since the open-system state can be mixed, we quantify this entanglement with the negativity $\mathcal{N}(\rho)=(\lVert\rho^{T_A}\rVert_1-1)/2$~\cite{zyczkowski1998volume,vidal2002negativity}. After tracing out the physical cavity, partial transposition of the emitter subsystem gives $\mathcal{N}=0.533$ at the final time. Its positive value certifies entanglement between the emitter and the two captured modes.

The preceding simulation can also be performed without the symbolic derivation, i.e., by directly using the numerical operators and functions. This is shown in \coderef{code:lambda_numeric}. We use the numerical operators and functions already defined in \coderef{code:lambda_translate_full} to create the \code{SLH} struct. The remaining procedure is unchanged.

\begin{lstlisting}[caption={The same two-mode simulation written directly with numerical operators.},label={code:lambda_numeric}]
# numeric Hamiltonian
Hₛ_slh = g_qo*(c_qo'*σ_qo(1, 3) + c_qo*σ_qo(3, 1) +
  c_qo'*σ_qo(2, 3) + c_qo*σ_qo(3, 2)) + ω₁₂_qo*σ_qo(2, 2)

# numeric SLH triples with time-dependent couplings
Gₛ_slh = SLH(1, √κ_qo*c_qo, Hₛ_slh)
Gᵥ₁_slh = SLH(1, t -> conj(gᵥ₁_qo(t))*aᵥ₁_qo, 0*one(b))
Gᵥ₂_slh = SLH(1, t -> conj(gᵥ₂_qo(t))*aᵥ₂_qo, 0*one(b))

# full Hamiltonian and jump operator
G_slh = cascade(Gₛ_slh, Gᵥ₁_slh, Gᵥ₂_slh)
H_slh = hamiltonian(G_slh)
L_slh = jump_operator(G_slh)[1]
\end{lstlisting}

\subsection{Quantum dots in a bidirectional waveguide with a coherent input pulse}
\label{sec:feedback}

As a second example, we consider two two-level emitters coupled to a bidirectional waveguide driven by a coherent input pulse~\cite{hansen2026waveguide}. Here the challenge is the network itself: the two propagation directions and the phase accumulated between the emitters generate coherent and dissipative couplings that depend on how the ports are connected. We construct the same physical network in two ways, first as explicit right- and left-moving cascades and then by closing the corresponding multiport network with feedback reduction.

We start by building the right- and left-moving channels explicitly and \code{concatenate} them. Emitter $i$ has detuning $\Delta_i$ and directional waveguide coupling rates $\gamma_R^{(i)}$ and $\gamma_L^{(i)}$, with $\hat{\sigma}_i^{mn}=\ket{m}_i\!\bra{n}_i$, and propagation between the emitters contributes the phase $\phi_{ij}$.
Unlike the preceding quantum-pulse examples, the input is represented by the coherent amplitude $E_{\rm in}(t)$ rather than a virtual input cavity. It enters through the right-moving channel, and the two output coupling operators describe transmission and reflection.
For $N=2$ emitters the symbolic construction reads:
\begin{lstlisting}[caption={Bidirectional waveguide constructed from cascades and concatenation.},label={code:waveguide_concat}]
using QuantumInputOutput
using SecondQuantizedAlgebra
using Symbolics: variable

# symbolic Hilbert space and operators
N = 2
hₐ(i) = NLevelSpace(Symbol("a$(i)"), 2)
h = tensor([hₐ(i) for i = 1:N]...)
σ(α, i, j) = Transition(h, Symbol("σ_$(α)"), i, j, α)

# symbolic parameters
γᵣ(i) = variable(Symbol("γ^{($(i))}_R"); T = Real)
γₗ(i) = variable(Symbol("γ^{($(i))}_L"); T = Real)
Δ(i) = variable(Symbol("Δ_{$(i)}"); T = Real)
ϕ(i, j) = variable(Symbol("ϕ_{$(i)$(j)}"); T = Real)
Eᵢₙ = variable(Symbol("E_{in}"); T = Real)

# drive, phase shift and QD SLH elements
Gᵢₙ = SLH(1, Eᵢₙ, 0)
G_ϕ(i, j) = SLH(exp(1im*ϕ(i, j)), 0, 0)
Gᵣ(i) = SLH(1, √γᵣ(i)*σ(i, 1, 2), -Δ(i)*σ(i, 2, 2))
Gₗ(i) = SLH(1, √γₗ(i)*σ(i, 1, 2), 0)

# cascade right- and left-moving modes
Gᵣ_tot = Gᵢₙ ▷ Gᵣ(1) ▷ G_ϕ(1, 2) ▷ Gᵣ(2)
Gₗ_tot = Gₗ(2) ▷ G_ϕ(1, 2) ▷ Gₗ(1)
# concatenate both channels
G_tot = Gᵣ_tot ⊞ Gₗ_tot

# full Hamiltonian and jump operators
H = hamiltonian(G_tot)
L = jump_operator(G_tot)
Lᵣ = L[1]
Lₗ = L[2]
\end{lstlisting}
The key structural point is that the right-moving propagation is built as a cascade in the physical propagation direction, while the left-moving propagation is constructed in reverse order. The
triple \code{G\_tot = Gᵣ\_tot} \codesymbol{$\boxplus$} \code{Gₗ\_tot}
then collects both channels into one two-port SLH model.
The resulting coupling vector contains one operator for each propagation direction.

The same physical model can also be constructed with the SLH \code{feedback} reduction rule (\secref{sec:slh}). Here one first concatenates the unconnected two-port components, keeps all waveguide links explicit as external ports, and then removes the internal connections symbolically with a sequence of feedback reductions. The feedback procedure applied in \coderef{code:feedback} corresponds to the one depicted in \fref{fig:SLH_rules}. 

\begin{lstlisting}[caption={Deriving the same effective model via concatenation and feedback reduction.},label={code:feedback}]
using LinearAlgebra

# individual SLH elements
I₂ = Matrix{Int}(I, 2, 2)
Gᵢₙ = SLH(I₂, [Eᵢₙ, 0], 0)
G_qd(i) = SLH(
  I₂, [√γᵣ(i)*σ(i, 1, 2), √γₗ(i)*σ(i, 1, 2)],
  -Δ(i)*σ(i, 2, 2))
G_ϕ(i, j) = SLH(
  [exp(1im*ϕ(i, j)) 0; 0 exp(1im*ϕ(i, j))], [0, 0], 0)

# feedback reduction
G_unc = Gᵢₙ ⊞ G_qd(1) ⊞ G_ϕ(1, 2) ⊞ G_qd(2)
G_red = feedback(G_unc,
  1 => 3, 3 => 5, 5 => 7, 8 => 6, 6 => 4, 4 => 2)

# full Hamiltonian and jump operators
H_red = hamiltonian(G_red)
L_red = jump_operator(G_red)
Lₗ_red = L_red[1]
Lᵣ_red = L_red[2]
\end{lstlisting}
The port numbers in \fref{fig:SLH_rules} and \coderef{code:feedback} follow from the concatenation. Each component has two ports, the first for the right-moving and the second for the left-moving channel. A pair \code{x => y} feeds output $x$ back into input $y$ [\eqsref{eq:fbS}--(\ref{eq:fbH})]. The right-moving links are therefore \code{1 => 3}, \code{3 => 5}, \code{5 => 7}, while the left-moving links run backwards, \code{8 => 6}, \code{6 => 4}, \code{4 => 2}. All numbers refer to the unreduced network \code{G\_unc}, and \code{feedback} keeps track of how the remaining ports are renumbered after each reduction. The ports left open are input 1, which carries the drive, output 7 (transmission), input 8, and output 2 (reflection). The open outputs keep their original order, so \code{L\_red[1]} is the reflected and \code{L\_red[2]} the transmitted field, opposite to \coderef{code:waveguide_concat}.

Feedback reduction is useful for networks whose internal connections are not naturally expressed as independent cascades. For this network, it gives the same Hamiltonian and coupling operators as the direct construction.
For two emitters with separation phase $\phi$, the resulting Hamiltonian is
\begin{align}
\hat{H}_{\rm wg}=&-\sum_{i=1}^2\Delta_i\hat{\sigma}_i^{22}
+\frac{1}{2i}\Big[E_{\rm in}\Big(\sqrt{\gamma_R^{(1)}}\hat{\sigma}_1^{21}
\nonumber\\
&\qquad+e^{i\phi}\sqrt{\gamma_R^{(2)}}\hat{\sigma}_2^{21}\Big)-\mathrm{h.c.} \Big]
\nonumber\\
&+\frac{1}{2i} \Big[e^{i\phi}\Big(\sqrt{\gamma_R^{(1)}\gamma_R^{(2)}}\hat{\sigma}_2^{21}\hat{\sigma}_1^{12}
\nonumber\\ &\qquad+\sqrt{\gamma_L^{(1)}\gamma_L^{(2)}}\hat{\sigma}_1^{21}\hat{\sigma}_2^{12}\Big)-\mathrm{h.c.} \Big].
\label{eq:waveguideH}
\end{align} 
The first bracket contains the coherent input pulse arriving at the two emitters with the correct propagation phase, while the second bracket contains the direction-dependent dipole--dipole exchange generated by the waveguide~\cite{Ostermann2019, Holzinger2022, Fasser2024, Tiranov2023,hansen2026waveguide}. This illustrates that the SLH network rules used here are not merely bookkeeping devices; they are also a compact way to derive the effective Hamiltonian and Lindblad operators for one-dimensional quantum systems.

\begin{figure}
\centering
\includegraphics[width=0.99\columnwidth]{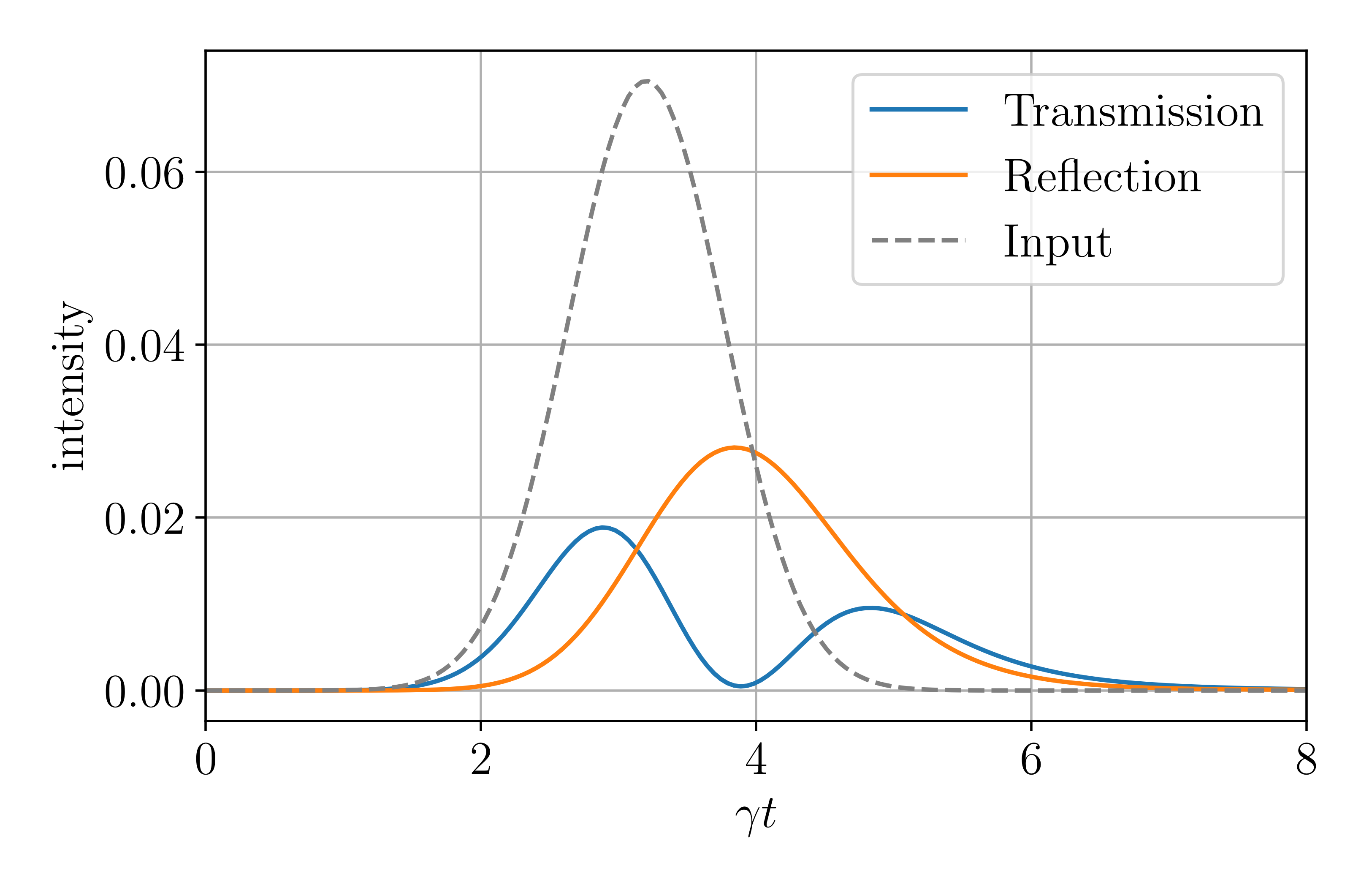}
\caption{Transmitted and reflected intensities for the
bidirectional
waveguide with two quantum dots driven by a coherent input pulse from the left. The dashed curve shows the
input intensity $|E_{\rm in}(t)|^2$.}
\label{fig:waveguide_intensity}
\end{figure}

For the numerical simulation we choose a Gaussian input envelope, translate the symbolic operators to \texttt{QuantumOptics.jl}, and compute the transmitted and reflected intensities.
\begin{lstlisting}[caption={Translation and numerical simulation of the output intensities for the bidirectional quantum-dot waveguide system.},label={code:feedback_translate}]
using QuantumOptics

# numeric parameters
γ = 1.0
β = 0.9
γᵣ_qo = fill(γ*β/2, N)
γₗ_qo = fill(γ*β/2, N)
γ_add = fill(γ*(1 - β), N)
Δ_qo = fill(0.0, N)
ϕ_qo = fill(pi/10, max(N - 1, 0))

# Gaussian input pulse
σₜ = 0.8
α₀ = √0.1
t₀ = 4σₜ
T_end = 3t₀
u(t) = 1/(√σₜ*pi^(1/4)) * exp(-(t - t₀)^2 / (2*σₜ^2))
Eᵢₙ_qo(t) = α₀*u(t)

# constant and time-dependent parameters
p = [[γᵣ(i) for i = 1:N]; [γₗ(i) for i = 1:N];
  [Δ(i) for i = 1:N]; [ϕ(i, i + 1) for i = 1:(N-1)]]
p_qo = [γᵣ_qo; γₗ_qo; Δ_qo; ϕ_qo]
parameter = Dict(p .=> p_qo)
time_parameter = Dict(Eᵢₙ => Eᵢₙ_qo)

# translate to numerics
bₐ = NLevelBasis(2)
b = tensor([bₐ for _ = 1:N]...)
H_qo = to_numeric(H, b; parameter, time_parameter)
Lᵣ_qo = to_numeric(Lᵣ, b; parameter, time_parameter)
Lₗ_qo = to_numeric(Lₗ, b; parameter, time_parameter)
# additional free-space decay
σ_qo(α, i, j) = to_numeric(σ(α, i, j), b)
L_add = [√γ_add[i]*σ_qo(i, 1, 2) for i = 1:N]

function f_io(t, ρ)
  Ht = H_qo(t)
  Lt = [Lᵣ_qo(t), Lₗ_qo(t), L_add...]
  return Ht, Lt, dagger.(Lt)
end

# time evolution
T = [0:0.005:1;] * T_end
ψ₀ = tensor([nlevelstate(bₐ, 1) for _ = 1:N]...)
t, ρt = timeevolution.master_dynamic(T, ψ₀, f_io)

# transmitted and reflected intensity
Iᵣ = zeros(length(t))
Iₗ = zeros(length(t))
for (i, ti) in enumerate(t)
  lᵣ = Lᵣ_qo(ti)
  lₗ = Lₗ_qo(ti)
  Iᵣ[i] = real(expect(lᵣ'*lᵣ, ρt[i]))
  Iₗ[i] = real(expect(lₗ'*lₗ, ρt[i]))
end
\end{lstlisting}
The resulting intensities are shown in \fref{fig:waveguide_intensity}. The dashed curve is the input pulse intensity $|E_{\rm in}(t)|^2$, while
the solid curves show the reflected field and the corresponding reshaping of the transmitted pulse by the two-emitter system.
In \appref{sec:qrt_waveguide}, we additionally provide a code sample to simulate the two-time intensity correlation
functions
in transmission and reflection.

\subsection{Interaction-picture scattering with a quantum pulse}
\label{sec:interaction_picture_example}

As a third example, we consider pulse scattering from a resonant two-level system in a regime where the direct virtual-cavity representation becomes inefficient. For a highly occupied pulse, the laboratory-frame basis must resolve the transfer of the full occupation from the input cavity to the output cavity, even though the scatterer changes only a small number of excitations. The interaction picture of \secref{sec:interaction_picture} removes these direct mode-transfer dynamics, so the truncated basis need only resolve the photons modified by the scattering. We take the input mode $u$ to be prepared in a Fock state with $n=100$ photons and choose the selected virtual output mode to have the same Gaussian profile, $v=u$.

We build the cascaded triple $G_u\triangleright G_s\triangleright G_v$, subtract the direct virtual-cavity Hamiltonian $\hat{H}_{uv}$, and substitute the transformed mode operators obtained from the matrix solution $M(t)$ of the interaction-picture evolution with \code{substitute}. This symbolic construction is shown in \coderef{code:interaction_pulse_example}.
The numerical translation to \texttt{QuantumOptics.jl} is shown separately in \coderef{code:interaction_pulse_example_num}, where we initialize the state $\ket{100}_u\ket{g}\ket{0}_v$, solve the master equation, and evaluate the
interaction-picture $u$- and $v$-mode occupations $\langle \hat{n}_u \rangle$ and $\langle \hat{n}_v \rangle$,
and the atomic excitation $\langle \hat{\sigma}^{22}\rangle$.

\begin{lstlisting}[caption={Symbolic interaction-picture construction for an $n=100$ Fock-state pulse scattering on a two-level system.},label={code:interaction_pulse_example}]
using QuantumInputOutput
using SecondQuantizedAlgebra
using Symbolics: variable

# symbolic Hilbert space and operators
h = FockSpace(:u) ⊗ NLevelSpace(:s, 2) ⊗ FockSpace(:v)

aᵤ = Destroy(h, :aᵤ, 1)
aᵥ = Destroy(h, :aᵥ, 3)
σ(i, j) = Transition(h, :σ, i, j, 2)
@variables gᵤ::Number gᵥ::Number γ::Real

# SLH elements, Hamiltonian and jump operator
Gᵤ = SLH(1, gᵤ'*aᵤ, 0)
Gₛ = SLH(1, √γ*σ(1, 2), 0)
Gᵥ = SLH(1, gᵥ'*aᵥ, 0)
G_cas = Gᵤ ▷ Gₛ ▷ Gᵥ
H = hamiltonian(G_cas)
L = jump_operator(G_cas)[1]

# interaction picture transformation
M(i, j) = variable(Symbol("M_{$(i)$(j)}");
  T = Complex{Real})
a₀ = [aᵤ, aᵥ]
a_int = [sum(M(i, j)*a₀[j] for j = 1:2) for i = 1:2]
int_dict = Dict(a₀ .=> a_int)
Hᵤᵥ = hamiltonian(Gᵤ ▷ Gᵥ)
H_int = substitute(H - Hᵤᵥ, int_dict)
L_int = substitute(L, int_dict)
\end{lstlisting}

We choose the parameters such that the atom reaches
approximately $50\%$ excitation,
using a Gaussian pulse $u(t)\propto\exp[-(t-t_{\rm p})^2/2\sigma_{\rm p}^2]$ centered at $t_{\rm p}=4\sigma_{\rm p}$ with a width of $\sigma_{\rm p} = \theta^2/( 8 n \gamma \sqrt{\pi} )$, where $\theta = \pi/2$ is the pulse area for the Rabi frequency $\Omega(t)=2\sqrt{\gamma n}\,u(t)$. In that case the atom and the pulse mode become entangled.

\begin{figure}
\centering
\includegraphics[width=0.99\columnwidth]{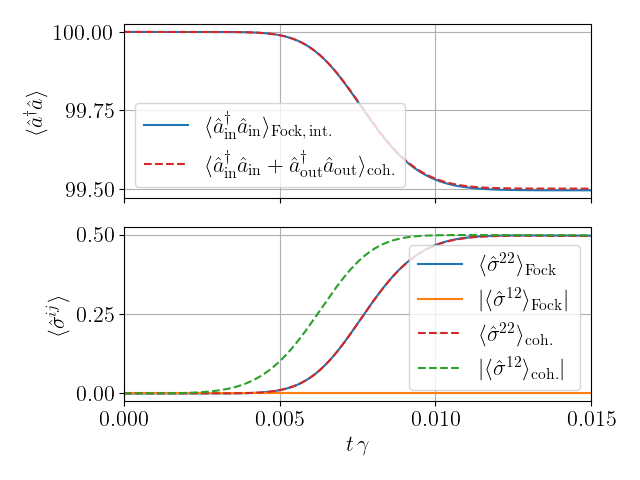}
\caption{Interaction picture and cumulant expansion. The top panel shows the mean photon number of the traveling light pulse in the temporal input mode for an initial Fock state and a coherent state, simulated with the interaction-picture master equation and a second-order cumulant expansion approach, respectively. 
The bottom panel shows the excited-state population and coherence. For the initial Fock state, the two-level system coherence vanishes for all times, as imposed by the phase symmetry of the master equation.}
\label{fig:interaction_picture_compare}
\end{figure}

In \fref{fig:interaction_picture_compare}, we plot the dynamics of the mean photon number (top) and the two-level system (bottom). The interaction-picture $u$-mode occupation
(blue solid line) is reduced by approximately one half of a photon after the interaction. The coherence of the two-level system vanishes at all times ($|\langle \hat{\sigma}^{12} \rangle| = 0$), due to the phase symmetry of the master equation. The atom and the field are entangled, which we quantify by the negativity across the atom--field bipartition of $\mathcal{N}=0.497$ at the end of the pulse, close to its maximum value of $1/2$ for an effective two-level bipartition. 

\begin{lstlisting}[caption={Numerical interaction-picture simulation for an $n=100$ Fock-state pulse.},label={code:interaction_pulse_example_num}]
using QuantumOptics

# numerical parameters
γ_qo = 1.0
nₚₕ = 100
σₚ = (π/2)^2 / (8*nₚₕ*γ_qo*√π)
tₚ = 4σₚ
T = [0:0.005:1;] * 0.015
u(t) = 1/(√σₚ*π^(1/4)) * exp(-0.5*((t - tₚ)/σₚ)^2)
gᵤ_qo = coupling_input(u, T)
gᵥ_qo = coupling_output(u, T)

# numerical interaction-picture coupling matrix
Aᵤᵥ = coupling_matrix((gᵤ_qo, gᵥ_qo))
M_qo = solve_mode_evolution(Aᵤᵥ, T)

parameter = Dict(γ => γ_qo)
Ms = [M(i, j) for i = 1:2 for j = 1:2]
Ms_qo = [t -> M_qo(t)[i, j] for i = 1:2 for j = 1:2]
time_parameter =
  Dict([gᵤ, gᵥ, Ms...] .=> [gᵤ_qo, gᵥ_qo, Ms_qo...])

# operator basis and translation to numerics
bᵤ = FockBasis(nₚₕ, nₚₕ - 2)
bₐ = NLevelBasis(2)
bᵥ = FockBasis(2)
b = bᵤ ⊗ bₐ ⊗ bᵥ
H_qo = to_numeric(H_int, b; parameter, time_parameter)
L_qo = to_numeric(L_int, b; parameter, time_parameter)
# time evolution
function f_io(t, ρ)
  Ht = H_qo(t)
  Lt = [L_qo(t)]
  return Ht, Lt, dagger.(Lt)
end
ψ₀ = fockstate(bᵤ, nₚₕ) ⊗ nlevelstate(bₐ, 1) ⊗
  fockstate(bᵥ, 0)
_, ρt = timeevolution.master_dynamic(T, ψ₀, f_io)

# expectation values
aᵤ_qo = destroy(bᵤ) ⊗ one(bₐ) ⊗ one(bᵥ)
aᵥ_qo = one(bᵤ) ⊗ one(bₐ) ⊗ destroy(bᵥ)
σee_qo = one(bᵤ) ⊗ transition(bₐ, 2, 2) ⊗ one(bᵥ)
σge_qo = one(bᵤ) ⊗ transition(bₐ, 1, 2) ⊗ one(bᵥ)

nᵤ = real.(expect(aᵤ_qo'*aᵤ_qo, ρt))
nᵥ = real.(expect(aᵥ_qo'*aᵥ_qo, ρt))
Pₑ = real.(expect(σee_qo, ρt))
s_ge = abs.(expect(σge_qo, ρt))

# entanglement between the atom and both field modes
N_atom_field = negativity(ρt[end], 2)
\end{lstlisting}

The symbolic SLH network constructed above can also be used without a Fock-space representation of the field. To illustrate this, we keep the same network but replace the Fock input by a coherent state with the same mean photon number $\bar n = |\alpha|^2 = 100$ and use a second-order cumulant expansion \cite{Plankensteiner2022quantumcumulantsjl}. 
A cumulant expansion truncates the hierarchy of operator averages, so the Fock-space cutoff of the previous example plays no role and we can stay in the laboratory frame. The interaction-picture transformation could be applied here as well.
We keep the symbolic SLH model, derive the second-order \code{meanfield} equations, \code{complete} the set of equations, and then integrate the resulting set of ODEs numerically.
There is one subtle derivation difference compared to the previous example. Since we do not translate the symbolic expressions to numerical operators, we must register the time-dependent pulse couplings as symbolic functions of the independent variable of the mean-field equations. To avoid repeating the SLH derivation, we \code{substitute} these registered functions directly into the already derived Hamiltonian and Lindblad operator. To obtain the initial average values for the time evolution we define the state first in \code{QuantumOptics.jl} as a \code{LazyKet} to keep the state vector small and then calculate the values with the function \code{initial\_values}.

\begin{lstlisting}[caption={Second-order cumulant treatment for a coherent pulse with mean photon number $\bar n=100$.},label={code:interaction_pulse_cumulants}]
using QuantumCumulants
using ModelingToolkitBase
using OrdinaryDiffEq

# time variable of the mean-field equations
@independent_variables t

# register the pulse couplings as functions of time
@register_symbolic gᵤ_reg(tt)
@register_symbolic gᵥ_reg(tt)
subs = Dict(gᵤ => gᵤ_reg(t), gᵥ => gᵥ_reg(t))
H_mf = substitute(H, subs)
L_mf = substitute(L, subs)
gᵤ_reg(tt) = gᵤ_qo(tt)
gᵥ_reg(tt) = gᵥ_qo(tt)

# derive and complete the mean-field equations
ops = [aᵤ, aᵤ'*aᵤ, σ(2, 2), aᵤ'*σ(1, 2),
  σ(1, 2), aᵥ, aᵥ'*aᵥ]
eqs = meanfield(ops, H_mf, [L_mf]; order = 2)
eqs_comp = complete(eqs)
sys = mtkcompile(System(eqs_comp; name = :sys))

# initial state
α₀ = √nₚₕ
bᵤ₂ = FockBasis(2*nₚₕ)
bₐ₂ = NLevelBasis(2)
bᵥ₂ = FockBasis(1)
b₂ = bᵤ₂ ⊗ bₐ₂ ⊗ bᵥ₂
ψ₀_coh = LazyKet(b₂, (coherentstate(bᵤ₂, α₀),
  nlevelstate(bₐ₂, 1), fockstate(bᵥ₂, 0)))

# define and solve the ODE problem
p₀ = merge(Dict(γ => γ_qo),
  initial_values(eqs_comp, ψ₀_coh))
prob = ODEProblem(sys, p₀, (0.0, T[end]))
sol = solve(prob, Tsit5(); saveat = T,
  abstol = 1e-10, reltol = 1e-10)

# expectation values
nᵤ₂ = real.(get_solution(sol, aᵤ'*aᵤ, eqs_comp).(T))
nᵥ₂ = real.(get_solution(sol, aᵥ'*aᵥ, eqs_comp).(T))
Pₑ₂ = real.(get_solution(sol, σ(2, 2), eqs_comp).(T))
s_ge₂ = abs.(get_solution(sol, σ(1, 2), eqs_comp).(T))
\end{lstlisting}
For example, for the excited-state population of the two-level system, the above code provides the mean-field equation
\begin{align}
    \partial_t {\langle \hat{\sigma}^{22} \rangle} = - \gamma \langle \hat{\sigma}^{22} \rangle - g_u(t) \sqrt{\gamma} \Big( \langle \hat{a}_u^\dagger \hat{\sigma}^{12} \rangle + \langle \hat{a}_u \hat{\sigma}^{21} \rangle \Big),
    \label{eq:popu_tls}
\end{align} 
where we used that the mode functions, and hence the virtual-cavity couplings, are real for the Gaussian pulse considered here. 
The dashed lines in \fref{fig:interaction_picture_compare} show the dynamics for the initial coherent state. We see that the mean photon number in the pulse is also reduced by approximately one half of a photon. Note, however, that because we are not operating in the interaction picture in this case, the change in the pulse is given by the sum of the input- and output-pulse virtual-cavity occupations, where we choose the temporal output mode to be equal to the input mode.
The crucial difference for the coherent-state pulse is that the atom retains coherence after the interaction. An additional exact master-equation calculation gives an atom--field negativity of only $\mathcal{N}=0.0135$, so the coherent pulse acts almost as a classical drive. This value cannot be obtained from the second-order cumulant averages alone.

\section{Conclusion and outlook}
\texttt{QuantumInputOutput.jl} combines SLH network composition with the virtual-cavity description of propagating quantum pulses in a common computational framework. Together, these tools allow network connectivity to determine the effective open-system model, relevant temporal output modes to be retained as explicit quantum subsystems, and the resulting dynamics to be treated in representations adapted to the problem. 

The virtual-cavity approach is most natural when the propagating field is concentrated in a small number of relevant temporal modes. For broad multimode fields or strong temporal entanglement, approaches that represent the propagating field explicitly, including time-bin and matrix-product-state methods~\cite{bundgaard2025waveguideqed,regidor2026qwavemps}, provide a complementary description.

Natural extensions include interfaces to other packages solving master equations~\cite{QuantumToolbox2025, Fishman2022}. Since the virtual-cavity construction yields a standard Lindblad master equation, established approximations~\cite{reiter2012effective, Plankensteiner2022quantumcumulantsjl} and numerical methods~\cite{molmer1992mcwf, ritsch1992mcwf, white1992mps, Verstraete2008, Ors2014} can be applied directly. An interface for \texttt{QuantumToolbox.jl}~\cite{QuantumToolbox2025} through \texttt{SecondQuantizedAlgebra.jl} has already been created.

We expect the framework to be useful for direct numerical simulations, but also for analytic derivations of input--output network equations. 

\section*{Acknowledgments}
The authors thank J.\ Kerber, H.\ Ritsch and Z.\ Aqua for testing the framework. C.H.\ was supported by the Carlsberg Foundation through the ``Semper Ardens'' Research Project QCooL. This research was funded in whole or in part by the Austrian Science Fund (FWF) 10.55776/J4865. For the purpose of open access, the authors have applied a CC BY public copyright license to any Author Accepted Manuscript (AAM) version arising from this submission. The source code of \texttt{QuantumInputOutput.jl} used in this article is archived on Zenodo~\cite{zenodo}.

\section*{Author contributions}
C.H.\ and O.A.\ developed the software package. K.M.\ supervised the work. All authors contributed to writing the manuscript.

AI tools (Claude by Anthropic and Codex by OpenAI) were used for the development of the package, including examples, documentation, tests, refactoring, and performance optimization. All output was reviewed, tested, and revised by the authors. AI tools were also used to draft and edit the text of the manuscript.

\clearpage
\newpage

\appendix

\section{Common SLH elements} \label{sec:SLH_elements}

In this section, we summarize the most frequently used SLH elements in \texttt{QuantumInputOutput.jl}, see also Appendix 1 of Ref.~\cite{combes2017slh}. Throughout, $\hat{\sigma}^{mn}=\ket{m}\!\bra{n}$ denotes a transition operator, with the level labels given for each element.

\begin{itemize}
\item Input (output) cavity for a traveling pulse with temporal input mode $u(t)$ (output mode $v(t)$):
\begin{align}
G_{u}=\left(1,\, g_u^*(t)\hat{a}_u,\,0\right)
\end{align}
\begin{align}
G_{v}=\left(1,\, g_v^*(t)\hat{a}_v,\,0\right)
\end{align}
\item Single-port phase shifter with phase $\phi$:
\begin{align}
G_{\mathrm{PS}}=\left(e^{i\phi},\,0,\,0\right)
\end{align}
\item Two-port beam splitter:
\begin{align} 
G_{\mathrm{BS}}=\left(
\begin{bmatrix}
r_{11} & t_{12}\\
t_{21} & r_{22}
\end{bmatrix},
\,0,\,
0
\right)
\end{align}

\item Coherent drive that displaces the input field by a generally time-dependent amplitude $\alpha(t)$:
\begin{align}
G_{\mathrm{coh}}=\left(1,\,\alpha(t),\,0\right)
\end{align}

\item One-sided cavity with decay rate $\kappa$, detuning $\Delta$, and bosonic mode operator $\hat{c} $:
\begin{align}
G_{\mathrm{1s}}=\left(1,\,\sqrt{\kappa}\,\hat{c} ,\,-\Delta \hat{c} ^\dagger \hat{c} \right)
\end{align}

\item Two-sided cavity (Fabry--P\'erot resonator) with detuning $\Delta$ and mirror decay rates $\kappa_1$ and $\kappa_2$:
\begin{align}
G_{\mathrm{2s}}=\left(
I_2,\,
\begin{bmatrix}
\sqrt{\kappa_1}\,\hat{c}\\
\sqrt{\kappa_2}\,\hat{c}
\end{bmatrix},
\,-\Delta \hat{c}^\dagger \hat{c}
\right)
\end{align}

\item Two-level system with ground state $| g \rangle$, excited state $| e \rangle$, and coupling to a chiral waveguide with guided decay rate $\gamma_g$, unguided decay rate $\gamma_\perp$, and transition frequency $\omega$:
\begin{align}
G_{\mathrm{TLS}}=\left(
I_2,\,
\begin{bmatrix}
\sqrt{\gamma_g}\,\hat{\sigma}^{ge}\\
\sqrt{\gamma_\perp}\,\hat{\sigma}^{ge}
\end{bmatrix},
\,\omega\hat{\sigma}^{ee}
\right) 
\end{align}

\end{itemize}

\section{Quantum regression theorem for the bidirectional waveguide}
\label{sec:qrt_waveguide}

In this section, we compute the two-time intensity correlations
\begin{align}
G^{(2)}(t_1,t_2)=\langle \hat{L}^\dagger(t_1)\hat{L}^\dagger(t_2)\hat{L}(t_2)\hat{L}(t_1)\rangle,
\label{eq:G2}
\end{align}
via the quantum regression theorem for the coherent-pulse waveguide example in \secref{sec:feedback}, with $\hat{L}$ the coupling operator of the transmitted or the reflected channel. For each time point $t_1$, one propagates the conditional state $\hat{L}(t_1)\rho(t_1)\hat{L}^\dagger(t_1)$ forward in time and evaluates the corresponding output intensity at later times $t_2$.

The function \code{correlation\_matrix}, which provides $g^{(1)}$ of \eqref{eq:g1} in \coderef{code:correlation}, has the same structure. For each $t_1$, it acts with $\hat{L}_s(t_1)$ on $\rho(t_1)$ from one side, propagates this operator forward in time, and evaluates $\hat{L}_s^\dagger$ at the later times $t_2$. The two calculations thus differ only in the conditional operator that starts the second propagation and in the operator measured at $t_2$. Since $G^{(2)}$ requires the two-sided conditional state $\hat{L}\rho\hat{L}^\dagger$, we write out the loop explicitly in \coderef{code:waveguide_qrt}.

\begin{lstlisting}[caption={Second-order output correlations from the quantum regression theorem.},label={code:waveguide_qrt}]
# time grid
nT = length(T)
G² = zeros(nT, nT)
G²_ref = zeros(nT, nT)

# output field operators, materialized at each time
L₀(t) = dense(static_operator(Lᵣ_qo(t)))
L₀_dag(t) = dagger(L₀(t))
L₀_ref(t) = dense(static_operator(Lₗ_qo(t)))
L₀_ref_dag(t) = dagger(L₀_ref(t))

for it₁ = 1:(nT-1)
  ρt₁ = ρt[it₁]

  # second time-evolution for transmission
  t₂, ρ₂ =
    timeevolution.master_dynamic(T[it₁:end],
      L₀(T[it₁]) * ρt₁ * L₀_dag(T[it₁]), f_io)

  # transmission two-time correlation
  G²_row = real.([
    expect(L₀_dag(t₂[j])*L₀(t₂[j]),
      ρ₂[j]) for j = 1:length(t₂)
  ])
  G²[it₁, it₁:end] = G²_row
  G²[it₁:end, it₁] = G²_row

  # second time-evolution for reflection
  t₂_ref, ρ₂_ref =
    timeevolution.master_dynamic(T[it₁:end],
      L₀_ref(T[it₁]) * ρt₁ * L₀_ref_dag(T[it₁]), f_io)

  # reflection two-time correlation
  G²_row_ref = real.([
    expect(L₀_ref_dag(t₂_ref[j])*L₀_ref(t₂_ref[j]),
      ρ₂_ref[j]) for j = 1:length(t₂_ref)
  ])
  G²_ref[it₁, it₁:end] = G²_row_ref
  G²_ref[it₁:end, it₁] = G²_row_ref
end
\end{lstlisting}

\begin{figure}
\centering
\includegraphics[width=\columnwidth]{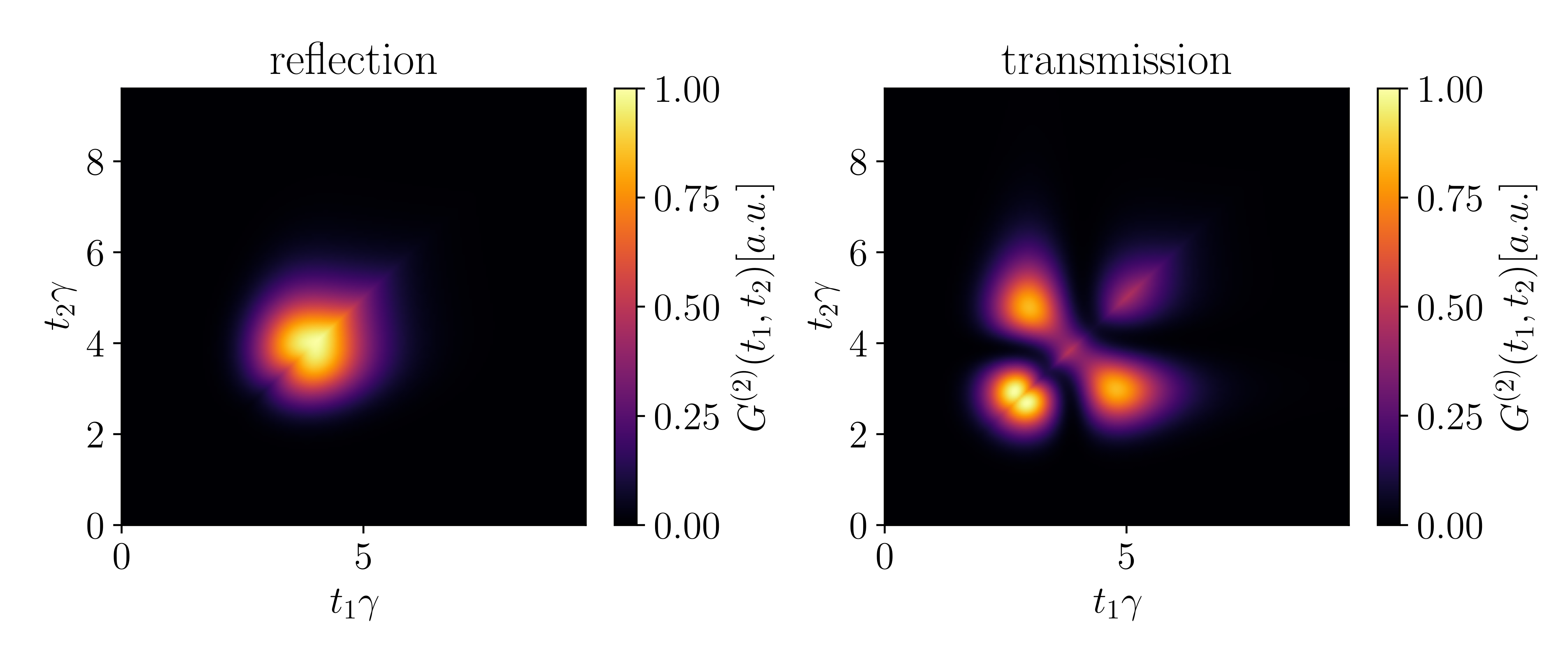}
\caption{Second-order correlation functions $G^{(2)}(t_1,t_2)$ for the reflected and transmitted fields of the bidirectional waveguide example, obtained via the quantum regression theorem. Each panel is scaled by its own maximum for display.}
\label{fig:waveguide_g2}
\end{figure}

The resulting correlation functions for the reflection and transmission are shown in \fref{fig:waveguide_g2}. In transmission, we see the characteristic bird shape~\cite{LeJeannic2022, regidor2026fewphoton}.

\section{Software versions} \label{sec:versions}

All code samples in this paper run with Julia \code{1.13.2} and the package versions listed in Table~\ref{tab:versions}. 

\begin{table}
\caption{Julia and package versions used for the code samples.}
\label{tab:versions}
\begin{ruledtabular}
\begin{tabular}{ll}
Software & Version \\
\hline
Julia & 1.13.2 \\
QuantumInputOutput.jl & 0.6.0 \\
SecondQuantizedAlgebra.jl & 0.15.0 \\
QuantumCumulants.jl & 0.8.0 \\
QuantumOptics.jl & 1.2.10 \\
QuantumOpticsBase.jl & 0.5.16 \\
ModelingToolkitBase.jl & 1.80.0 \\
OrdinaryDiffEq.jl & 7.8.1 \\
Symbolics.jl & 7.48.0 \\
\end{tabular}
\end{ruledtabular}
\end{table}

\bibliography{QIORef}

\end{document}